\documentclass[12pt]{article}
\usepackage{color}
\usepackage{subfigure}
\usepackage{epsfig}
\usepackage{amssymb,amsmath}

\begin{document}

\begin{titlepage}
\begin{flushright}
HIP-2009-04/TH
\\
26 March, 2009\\
\end{flushright}
\vfill
\begin{centering}

{\bf EPS09 --- A NEW GENERATION OF NLO AND LO NUCLEAR PARTON DISTRIBUTION FUNCTIONS}

\vspace{0.5cm}
K.~J. Eskola$^{\rm a,b,}$\footnote{kari.eskola@phys.jyu.fi},
H. Paukkunen$^{\rm a,b,}$\footnote{hannu.paukkunen@phys.jyu.fi} and 
C.~A. Salgado$^{\rm c,}$\footnote{carlos.salgado@usc.es}

\vspace{1cm}
{\em $^{\rm a}$Department of Physics,
P.O. Box 35, FI-40014 University of Jyv\"askyl\"a, Finland}
\vspace{0.3cm}

{\em $^{\rm b}$Helsinki Institute of Physics,
P.O. Box 64, FI-00014 University of Helsinki, Finland}
\vspace{0.3cm}

{\em $^{\rm c}$Departamento de F\'\i sica de Part\'\i culas and IGFAE, Universidade de Santiago de 
Compostela, Spain}

\vspace{1cm} 
{\bf Abstract} \\ 
\end{centering}
We present a next-to-leading order (NLO) global DGLAP analysis of nuclear parton distribution functions (nPDFs) and their uncertainties. Carrying out an NLO nPDF analysis for the first time with three different types of experimental input --- deep inelastic $\ell$+A scattering, Drell-Yan dilepton production in p+$A$ collisions, and inclusive pion production in d+Au and p+p collisions at RHIC --- we find that these data can well be described in a conventional collinear factorization framework. Although the pion production has not been traditionally included in the global analyses, we find that the shape of the nuclear modification factor $R_{\rm dAu}$ of the pion $p_T$-spectrum at midrapidity retains sensitivity to the gluon distributions, providing evidence for shadowing and EMC-effect in the nuclear gluons. We use the Hessian method to quantify the nPDF uncertainties which originate from the uncertainties in the data. In this method the sensitivity of $\chi^2$ to the variations of the fitting parameters is mapped out to orthogonal error sets which provide a user-friendly way to calculate how the nPDF uncertainties propagate to any factorizable nuclear cross-section. The obtained NLO and LO nPDFs and the corresponding error sets are collected in our new release called {\ttfamily EPS09}. These results should find applications in precision analyses of the signatures and properties of QCD matter at the LHC and RHIC.

\noindent

\vfill
\end{titlepage}

\setcounter{footnote}{0}

\section{Introduction}
\label{sec:intro}

In the advent of the LHC becoming online, the knowledge of free and bound nucleon parton distribution functions (PDFs) together with their uncertainties, is perhaps more topical than ever. Reliable PDF estimates play a significant role in extracting the dynamics of the underlying partonic processes from the plethora of detected events in a hadron collider like Tevatron or LHC.
 
Building on our experience obtained in a series of leading order (LO) analyses \cite{Eskola:1998iy,Eskola:1998df,Eskola:2007my,Eskola:2008ca} of the nuclear PDFs (nPDFs), we here extend our work to the next-to-leading order (NLO) accuracy in perturbative QCD (pQCD). The earlier studies on this topic \cite{Hirai:2007sx,deFlorian:2003qf} have already demonstrated that this non-trivial extension is successful, which strongly supports the collinear-factorization approach in describing the hard nuclear collisions. In addition to confirming this conclusion, our analysis  is an extension in two important ways:

First, in contrast to the earlier NLO nPDF analyses that were restricted to include only data from deep inelastic lepton-nucleus scattering (DIS) and Drell-Yan (DY) dilepton production, our analysis contains also inclusive pion production measured at RHIC. This additional input improves especially the determination of the gluon densities.

The second extension concerns the nPDF uncertainty analysis. Although this issue has been pursued earlier \cite{Hirai:2007sx,Hirai:2004wq,Eskola:2007my}, it has been very difficult for a general user to actually compute the propagation of  the nPDF uncertainties to a cross-section of his/her interest. The necessary tools for doing such estimates have only been available for the groups who did the nPDF analysis and in practice one has hitherto been forced to only compare the best-fit predictions given by independent nPDF fits, see e.g. Ref.~\cite{Cazaroto:2008qh}. To improve upon this issue we follow the ideas originally developed by the CTEQ collaboration \cite{Stump:2001gu,Pumplin:2001ct}, and construct a collection of nPDF error sets meant to serve as a practical way for estimating the nPDF-originating uncertainties in the hard process cross-sections in nuclear collisions. The package named {\ttfamily EPS09}, consisting of our best NLO and LO fits together with 30 error sets for both, will be available in \cite{EPS09code}.

In what follows, we will first introduce the general framework and the method of the present analysis in Sec.~\ref{sec:framework}. The obtained results are presented in Sec.~\ref{sec:results}, which also contains comparisons with the data and with earlier global analyses. In Sec.~\ref{sec:Application}, we present a concrete application of the obtained nPDFs and show how to best estimate the PDF-related uncertainties for nuclear cross-sections. In Sec.~\ref{sec:Summary} we give a summary and a brief outlook. The technical details of our error analysis are explained in Appendix~\ref{sec:ErrorAnalysis}, and in Appendix~\ref{InclusivePionProduction} we show how to speed up the numerical calculation for pion production.

\section{Framework and analysis method}
\label{sec:framework}

\subsection{Definition of nPDFs}
\label{subsec:nPDFs}

Following the framework of our earlier analyses, we define the bound proton NLO PDFs $f_{i}^A(x,Q^2)$ for each parton flavor $i$ by
\begin{equation}
f_{i}^A(x,Q^2) \equiv R_{i}^A(x,Q^2) f_{i}^{\rm CTEQ6.1M}(x,Q^2),
\end{equation}  
where $R_{i}^A(x,Q^2)$ denotes the nuclear modification to the free proton PDF $f_{i}^{\rm CTEQ6.1M}$ from the CTEQ6.1M set \cite{Stump:2003yu} in the $\overline{MS}$ scheme. Although more recent sets of free proton PDFs exist, this is the latest CTEQ set in the zero-mass variable flavour number scheme (ZM-VFNS) which we also use here --- a prescription that treats the heavy quarks as massless particles, each one active only when the factorization scale $Q^2$ exceeds its mass threshold $Q^2 \ge m_q^2$. We assume that isospin symmetry for bound protons and neutrons holds and take the average up and down quark PDF in a nucleus $A$ with $Z$ protons to be
\begin{eqnarray}
u_A(x,Q^2) & = & \frac{Z}{A} f_u^A(x,Q^2) + \frac{A-Z}{A} f_d^A(x,Q^2) \\
d_A(x,Q^2) & = & \frac{Z}{A} f_d^A(x,Q^2) + \frac{A-Z}{A} f_u^A(x,Q^2), \nonumber
\end{eqnarray}
like in our earlier works \cite{Eskola:1998iy,Eskola:1998df,Eskola:2007my,Eskola:2008ca}. We neglect the nuclear modifications in Deuterium $A=2$. Such effects were studied e.g. in Ref.~\cite{Hirai:2007sx} where an order of $1\ldots2\%$ effects were suggested. However, since only a minor improvement in $\chi^2$ was found, we expect that the error we make by neglecting the nuclear effects in Deuterium should be negligible compared to typical uncertainty originating from the experimental errors.

The parametrization of the nuclear modifications $R_{i}^A(x,Q^2)$ is performed at the charm quark mass threshold $Q_0^2 \equiv m_c^2=1.69 \, {\rm GeV}^2$ imposing the momentum and baryon number sum rules
\begin{equation}
 \sum_{i=q,\overline{q},g} \int_0^1 dx \, xf_{i}^A(x,Q_0^2) = 1, \quad \int_0^1 dx \left[ f_{u_V}^A(x,Q_0^2) + f_{d_V}^A(x,Q_0^2) \right] = 3, \label{eq:sumrules}
\end{equation}
for each nucleus $A$ separately. At higher scales $Q^2 > Q_0^2$, the nPDFs are obtained by solving the DGLAP evolution equations \cite{DGLAP} with NLO $\overline{MS}$ splitting functions \cite{Curci:1980uw,Furmanski:1980cm,Ellis:1991qj}.

We solve these equations numerically for both the nuclear and the free proton PDFs, following an efficient method suggested in \cite{Santorelli:1998yt}: The PDFs are approximated by 3th order polynomials in sufficiently small $x$-intervals for which the convolution integrals involving the splitting kernels can be evaluated analytically. The evolution in $Q^2$ is built as a Taylor expansion around the initial scale $Q_0^2$. We have tested the accuracy of our code against the benchmark PDFs \cite{Giele:2002hx}, and observed that 9th order Taylor expansion with $\sim 200$ $x$-points reaches an accuracy well below 1\% at $Q^2 = 10^4 \, {\rm GeV}^2$ in all $x$ except the largest-$x$ ($x \ge 0.9$) sea-quark sector. In practice, for the global fit procedure we keep the 9th order Taylor expansion but coarsen the $x$-grid to speed up the computation but ensure that the evolution works properly in the required kinematical domain $Q^2 < 200 \, {\rm GeV}^2$ (see Fig.~\ref{Fig:KinematicReach} ahead).

The relevant parameters that specify the evolution are the charm and bottom quark masses $m_c=1.3 \, {\rm GeV}$ and $m_b=4.5 \, {\rm GeV}$, and the 2-loop strong coupling constant fixed at $Z$-pole to $\alpha_s(M_Z^2)=0.118$. The top quark is not considered in the present analysis. These choices coincide with the ones adopted in CTEQ6.1M. In the LO analysis we take the free proton baseline from the CTEQ6L1 \cite{Pumplin:2002vw} set, and solve the LO DGLAP equations with 1-loop $\alpha_s$.

\subsection{Fitting functions and parameters}
\label{subsec:parameters}

The diversity of the presently available data is not broad enough to allow the nuclear modifications for each parton flavor to be independently determined. Thus, following our previous works \cite{Eskola:1998iy,Eskola:1998df,Eskola:2007my,Eskola:2008ca} we define only three different nuclear corrections at the initial scale $Q_0^2$: $R_V^A$ for both valence quark distributions, $R_S^A$ for all sea quarks, and $R_G^A$ for gluons. We parametrize these by piecewize functions
\begin{equation}
R^A_i(x) = \left\{
 \begin{array}{ll}
  a_0+(a_1+a_2 x)[\exp(-x) 
           - \exp(-x_a)]   &  x \, \, \le x_a \\
  b_0 + b_1 x + b_2 x^2 + b_3 x^3
             & x_a \le x \le x_e \\ 
  c_0 + (c_1-c_2x)(1-x)^{-\beta}    
             & x_e \le x \le 1,
\end{array}
\right.
\label{eq:parametrization}
\end{equation}
where the parameters $a_i$, $b_i$, $c_i$, $\beta$, $x_a$ and $x_e$ are $A$-dependent. Joining the three pieces together to give a continuous function with vanishing first derivatives at matching points $x_a$ and $x_e$, eliminates 6 out of the 13 parameters. The remaining ones are expressed in terms of the following 6 parameters with obvious interpretations: \\

\begin{figure}[htbp]
\center
\includegraphics[scale=0.7]{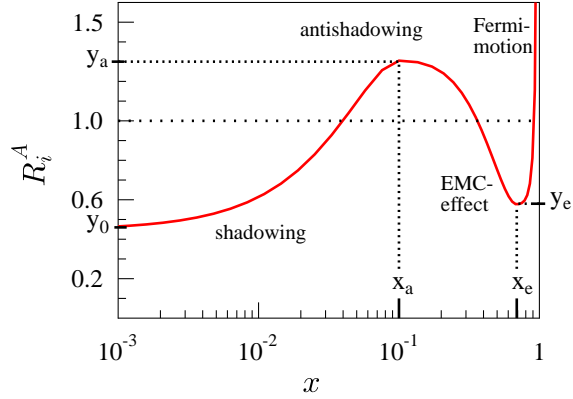}
\caption[]{\small An illustration of the fit function $R_i^A(x)$ and the role of the parameters $x_a$, $x_e$, $y_0$, $y_a$, and $y_e$.}
\label{Fig:Example}
\end{figure}

\begin{tabular}{ll}
  $y_0$            & Height to which shadowing levels as $x \rightarrow 0$ \\
  $x_a$, $y_a$     & Position and height of the antishadowing maximum \\
  $x_e$, $y_e$     & Position and height of the EMC minimum \\
  $\beta$        	 & Slope factor in the Fermi-motion part, \\
\end{tabular} 
\\

\noindent the remaining parameter $c_0$ is fixed to $c_0=2y_e$. The roles of these parameters are illustrated in Fig.~\ref{Fig:Example} which also roughly indicates which $x$-regions are meant by the commonly used terms: shadowing, antishadowing, EMC-effect, and Fermi-motion. 

The $A$-dependence of the fit parameters is assumed to follow a power law
\begin{equation}
  d_i^A = d_i^{A_{\rm ref}} \left( \frac{A}{A_{\rm ref}} \right)^{\,p_{d_i}},
  \label{eq:Adependence}
\end{equation}
where $d_i = x_a, y_a\ldots$, and where the reference nucleus is Carbon, $A_{\rm ref}=12$.

The baryon number and momentum sum rules eliminate $y_0$ and $p_{y_0}$ for valence quarks and gluons, leaving us with 32 free parameters. This is still way too large number of parameters to be determined only by the data --- further assumptions (based on prior experience) are needed to decide which parameters can truly be deduced from the data and which can be taken as fixed.

\subsection{Experimental input and cross-sections}
\label{subsec:ExperimentalInput}

\begin{table}
\begin{center}
{\footnotesize
\begin{tabular}{lclcccccc}
 Experiment & Process &  Nuclei & Data points & $\chi^2$ LO & $\chi^2$ NLO & Weight & Ref. \\
\hline
\hline
 SLAC E-139	& DIS	& He(4)/D & 21 & 6.5 & 7.3 &  1 & \cite{Gomez:1993ri} \\
 NMC 95, re. & DIS	& He/D  	& 16 & 14.5 & 15.6 & 5 & \cite{Amaudruz:1995tq} \\
 \\
 NMC 95         & DIS	& Li(6)/D	& 15 & 23.6 & 16.8 & 1 & \cite{Arneodo:1995cs} \\
 NMC 95, $Q^2$ dep.  & DIS	& Li(6)/D	& 153 & 162.2 & 157.0 &  1 & \cite{Arneodo:1995cs} \\
  \\
 SLAC E-139     	& DIS	& Be(9)/D & 20 & 9.6 & 12.2 & 1 & \cite{Gomez:1993ri}  \\
 NMC 96            & DIS	& Be(9)/C & 15 & 3.8  & 3.8 & 1 &  \cite{Arneodo:1996rv} \\
 \\
 SLAC E-139	    	& DIS	& C(12)/D &  7 & 4.1 & 3.2 & 1 &  \cite{Gomez:1993ri}  \\
 NMC 95    		      & DIS	& C/D     & 15 & 15.0 & 13.8 & 1 & \cite{Arneodo:1995cs} \\
 NMC 95, $Q^2$ dep. & DIS	& C/D     & 165 & 141.8 & 142.0 & 1 & \cite{Arneodo:1995cs}  \\ 
 NMC 95, re.  & DIS	& C/D   	& 16 & 19.3 & 20.5 & 1 & \cite{Amaudruz:1995tq} \\
 NMC 95, re.  & DIS	& C/Li   	& 20 & 30.3 & 28.4 & 1 & \cite{Amaudruz:1995tq} \\
 FNAL-E772 		      & DY	& C/D     &  9 & 7.5   & 8.3 & 1 & \cite{Alde:1990im}    \\
\\
 SLAC E-139		& DIS	& Al(27)/D      & 20 & 10.9 & 12.5 & 1 & \cite{Gomez:1993ri}   \\
 NMC 96    		& DIS 	& Al/C        & 15 & 6.0 & 5.8 & 1 & \cite{Arneodo:1996rv} \\
\\
 SLAC E-139		& DIS	& Ca(40)/D      &  7 & 5.0 & 4.1 & 1 & \cite{Gomez:1993ri}   \\
 FNAL-E772 	& DY	& Ca/D          &  9 & 2.9  & 3.4 &  15 & \cite{Alde:1990im}  \\
 NMC 95, re. 	& DIS	& Ca/D  & 15 & 25.4 & 24.7 & 1 & \cite{Amaudruz:1995tq} \\
 NMC 95, re. 	& DIS	& Ca/Li & 20 & 23.9 & 19.6 & 1 & \cite{Amaudruz:1995tq} \\
 NMC 96    		& DIS	& Ca/C          & 15 & 6.0  & 6.0 & 1 & \cite{Arneodo:1996rv} \\
\\
 SLAC E-139		& DIS	& Fe(56)/D      & 26 & 19.1 & 23.9 & 1 & \cite{Gomez:1993ri}   \\
 FNAL-E772 		& DY	& Fe/D          &  9 & 2.1 & 2.2 & 15 & \cite{Alde:1990im}    \\
 NMC 96    		& DIS	& Fe/C        & 15 & 11.0 & 10.8 & 1 & \cite{Arneodo:1996rv} \\
 FNAL-E866  & DY	& Fe/Be         & 28 & 20.9 & 21.7 & 1 &  \cite{Vasilev:1999fa} \\
\\ 
 CERN  EMC & DIS	& Cu(64)/D      & 19 & 13.4 & 14.8 & 1 & \cite{Ashman:1992kv}   \\
\\
 SLAC E-139	& DIS	& Ag(108)/D     &  7 & 3.8 & 2.9 & 1 & \cite{Gomez:1993ri}   \\
\\
 NMC 96  		& DIS  	& Sn(117)/C   & 15 & 9.6 & 9.1 & 1 & \cite{Arneodo:1996rv} \\
 
 NMC 96, $Q^2$ dep.  	& DIS	& Sn/C  & 144 & 80.2 & 82.8 & 10 & \cite{Arneodo:1996ru} \\
{} & {} & {} & {} & {} & {} & {} $\stackrel{(\rm x = 0.0125 \, only)}{}$  \\ 
 FNAL-E772 	& DY	& W(184)/D      &  9 & 7.0 & 6.7 & 10 & \cite{Alde:1990im}    \\
 FNAL-E866 	& DY	& W/Be          & 28 & 27.3 & 24.2 & 1 & \cite{Vasilev:1999fa} \\
\\
 SLAC E-139	& DIS	& Au(197)/D     & 21 & 11.6 & 13.8 & 1 & \cite{Gomez:1993ri}   \\
 RHIC-PHENIX  & $\pi^0$ prod.	& dAu/pp        & 20 & 7.3 & 6.3 & 20 & \cite{Adler:2006wg}\\

\\
 NMC 96 & DIS	& Pb/C          & 15 & 6.90 & 7.2 & 1 & \cite{Arneodo:1996rv}\\
\\

 \hline		   
 Total 		     &        &     &  929 & 738.6 &  731.3&                       \\
\end{tabular}
}
\caption[]{\small The data used in our analysis. The mass numbers are indicated in parentheses and the number of data points refers to those falling within our kinematical cuts, $Q^2, M^2 \ge 1.69 \, {\rm GeV}^2$ for DIS and DY, and $p_T \ge 1.7 \, {\rm GeV}$ for hadron production at RHIC. The quoted $\chi^2$ values correspond to the unweighted contributions of each data set in LO and NLO. Also the weight factors for data sets are shown.}
\label{Table:Data}
\end{center}
\end{table}

\begin{figure}[htbp]
\center
\includegraphics[scale=0.7]{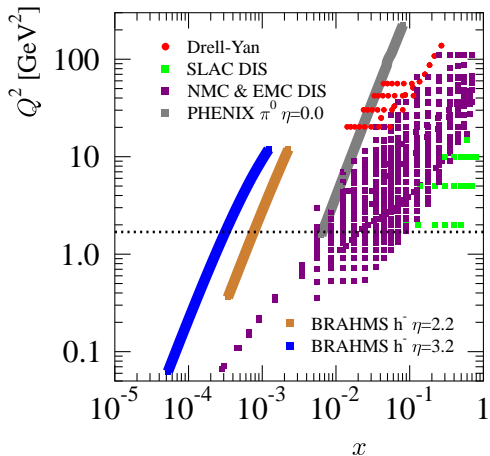}
\caption[]{\small The kinematical reach of the DIS, DY and pion production data (see Table \ref{Table:Data}) corresponding to the factorization scale choices explained in the text. The points indicate the lowest $x$ and $Q^2$ values in which partons are sampled in the cross-section calculation. Also the BRAHMS data \cite{Arsene:2004ux} for negatively charged hadron production is shown as it will be discussed later in Sec.~\ref{sec:Application}. The dashed horizontal line indicates the kinematical cut imposed on the data.}
\label{Fig:KinematicReach}
\end{figure}

The main body of the data in our analysis consists of $\ell+A$ DIS measurements. We also utilize the DY dilepton production data from fixed target p+$A$ collisions at Fermilab and inclusive neutral-pion production data measured in d+Au and p+p collisions at RHIC\footnote{In contrast to our previous analysis \cite{Eskola:2008ca}, we do not include the BRAHMS forward rapidity charged hadron d+Au data here. These data will be  separately discussed in Sec.~\ref{sec:Application}.}. Table~\ref{Table:Data} lists the sets included in our analysis and Fig.~\ref{Fig:KinematicReach} displays their kinematical reach in the $(x,Q^2)$-plane. We will use the following notation:
\begin{eqnarray}
R_{\rm DIS}^{\rm A}(x,Q^2)  & \equiv & \frac{\frac{1}{A}d\sigma_{\rm DIS}^{l \rm
{A}}/dQ^2dx}{\frac{1}{2}d\sigma_{\rm DIS}^{l{\mathrm d}}/dQ^2dx}, \qquad R_{F_2}^{\rm 
A}(x,Q^2) \equiv  \frac{F_2^A(x,Q^2)}{F_2^d(x,Q^2)} \nonumber \\
R_{\rm DY}^{\rm A}(x_{1,2},M^2) & \equiv & \frac{\frac{1}{A}d\sigma^{\rm pA}_{\rm 
DY}/dM^2dx_{1,2}}{\frac{1}{2}d\sigma^{\rm pd}_{\rm DY}/dM^2dx_{1,2}} \\
R_{\rm dAu}^{\pi} & \equiv & \frac{1}{\langle N_{\rm coll}\rangle} \frac{d^2 N_{\pi}^{\rm dAu}/dp_T dy}{d^2 N_{\pi}^{\rm pp}/dp_T dy} \stackrel{\rm min. bias}{=} \frac{\frac{1}{2A} d^2\sigma_{\pi}^{\rm dAu}/dp_T dy}{d^2\sigma_{\pi}^{\rm pp}/dp_T dy}. \nonumber 
\label{RF2RDY}
\end{eqnarray}
The kinematical variables in DIS are the Bjorken-$x$ and the virtuality of the photon $Q^2$. In DY $M^2$ denotes the invariant mass of the lepton pair, and $x_{1,2} \equiv \sqrt{M^2/s}\,e^{\pm y}$ where $y$ is the pair rapidity. The inclusive pion production is characterized by the transverse momentum $p_T$ and rapidity $y$ of the outgoing pion. The average number of binary nucleon-nucleon collisions (in the centrality class studied) is denoted by $\langle N_{\rm coll}\rangle$. In this analysis we only consider minimum bias data, and do not focus on the transverse coordinate dependence of the nPDFs. The kinematical cuts we impose on the data are $M^2, Q^2 \ge 1.69 \, {\rm GeV}^2$ for DIS and DY, and $p_T \ge 1.7 \, {\rm GeV}$ for inclusive pion production.

All cross-sections are calculated in the collinear factorization formalism folding the PDFs $f_i(x,\mu^2)$ with perturbatively calculable pieces $\hat{\sigma}$, schematically
\begin{eqnarray}
\sigma^{\ell + A \rightarrow \ell + X}_{\rm DIS} & = & \sum_{i=q,\overline{q},g} f_i^A(\mu^2) \otimes \hat{\sigma}_{\rm DIS}^{\ell + i \rightarrow \ell + X}(\mu^2)
\label{eq:factorized} \nonumber  \\
\sigma^{p + A \rightarrow l^+ l^- + X}_{\rm DY} & = & \sum_{i,j=q,\overline{q},g} f_i^p(\mu^2) \otimes f_j^A(\mu^2) \otimes \hat{\sigma}^{ij\rightarrow l^+ l^- + X}(\mu^2) \\
\sigma^{A+B \rightarrow \pi + X} & = & \sum_{i,j,k=q,\overline{q},g} f_i^A(\mu^2) \otimes f_j^B(\mu^2) \otimes \hat{\sigma}^{ij\rightarrow k + X}(\mu^2) \otimes D_{k \rightarrow \pi}(\mu^2). \nonumber 
\end{eqnarray}
As indicated above, we choose the factorization, renormalization and fragmentation scales to be equal, fixed to a characteristic scale in the process: in DIS to the photon virtuality $Q^2$, in DY to the invariant mass of the lepton pair $M^2$, and in the pion production to its transverse momentum $p_T^2$. 

The expressions for double differential DIS cross-sections $d^2\sigma/dxdQ^2$ involving the NLO coefficients for structure functions $F_{1,2}$ are well documented e.g. in Ref. \cite{Furmanski:1981cw}. In the case of DY we need the rapidity distributions $d^2\sigma/dM^2dy_R$, from which $d^2\sigma/dM^2dx_{1,2}$ are obtained by the relation $x_{1,2}=\sqrt{M^2/s}\,e^{\pm y}$. The relevant partonic cross-sections for DY process can be obtained from \cite{Anastasiou:2003yy}.

The numerical calculation of NLO DIS cross-sections involve only one-dimensional integrals, posing no problems in evaluating them repeatedly during the fitting process. The calculation of the DY cross-section, however, requires evaluation of double integrals which are already somewhat slower to compute. The small amount of available DY data nevertheless keeps the direct evaluation of the integrals still fast enough, and we do not need to resort to use pre-computed K-factors as is done e.g in \cite{Martin:2009iq}.

The inclusive pion production embodies the fragmentation functions $D_{i\rightarrow \pi}(z,\mu^2)$ for parton flavour $i$ to make a pion carrying a fraction $z$ of its momentum. In this work we have decided to use the KKP average pion $\pi^++\pi^-$ parametrization \cite{Kniehl:2000fe} for describing the neutral pion $\pi^0$ production. We have, however, checked that the results obtained with newer ones AKK08 \cite{Albino:2008fy} and nDSS \cite{de Florian:2007hc,deFlorian:2007aj} practically all coincide in the case of average pions. In practice, we employ the {\ttfamily INCNLO} \cite{INCNLO} program which includes the NLO hadron production code by Aversa et al. \cite{Aversa:1988vb}. The computation of pion production in NLO involves triple integrals which would be very time consuming to evaluate directly at every round of iteration, but we can speed up the computation significantly by applying the procedure explained in Appendix~\ref{InclusivePionProduction}.

The role played by each data type can be summarized as follows:

\begin{itemize}
\item {\bf Deep inelastic scattering} \\
DIS data has an excellent constraining power for quark distributions in the whole range $0.01 \le x \le 1$ spanned by the data. At large $x$ these data are mainly sensitive to the valence quarks and at small $x$ to the sea quarks. At moderate $x$, however, close to the antishadowing peak near $x=0.1$, such separation of the sea and valence quark contributions is not possible on the basis of this type of data alone. Despite the direct photon-gluon fusion channel contributing to the DIS cross-section at NLO, the main gluon constraint provided by DIS still comes through the scale evolution of sea quarks that is driven by the gluons.

\item {\bf Drell-Yan dilepton production} \\
The DY data, taken together with DIS, can discriminate between valence and sea quarks near $x=0.1$. The DY cross-section retains also some sensitivity to the sea quarks at larger $x$ but, unfortunately, the precision of the current data is not enough to exploit this constraint in its full potential. The invariant mass $M^2$ in our data sample is typically large, $M^2 \gg Q_0^2$, and consequently there are sizable evolution effects that constrain the gluons also.

\item {\bf Inclusive pion production} \\
This type of data is usually accompanied by a rather large normalization uncertainty stemming, among other sources, from the model-dependent quantity $\langle N_{\rm coll}\rangle$. Apart from the normalization uncertainty, the \emph{shape} of $R_{\rm dAu}^\pi$ can nevertheless act as a vital constraint, especially for the nuclear modification for gluons. The slight downward trend seen in the large-$p_T$ part of $R_{\rm dAu}^\pi$  at midrapidity \cite{Adler:2006wg} indicates the need for a gluon EMC-effect, and the smallest-$p_T$ part of the $R_{\rm dAu}^\pi$ would not be satisfactorily reproduced without shadowing (see Fig.~\ref{Fig:PHENIX} ahead).
 
\end{itemize}

\subsection{Definition of the best fit}
\label{sec:ChiSquared}

A customary way of finding an optimal set of parameters that fits a large number of experimental data, is to minimize a global $\chi^2$-function with the help of a routine such as {\ttfamily MINUIT} \cite{James:1975dr}. The requirement for a $\chi^2$-function is that its magnitude measures the degree of agreement between the theory predictions and the experimental data, but the exact definition can vary from one analysis to another. In this work we define the global $\chi^2$ by \cite{Stump:2001gu,Eskola:2008ca}
\begin{eqnarray}
\chi^2(\{a\})   & \equiv & \sum _N w_N \, \chi^2_N(\{a\}) 
\label{eq:chi2mod_1}
\\	
\chi^2_N(\{a\}) & \equiv & \left( \frac{1-f_N}{\sigma_N^{\rm norm}} \right)^2 + \sum_{i \in N}
\left[\frac{ f_N D_i - T_i(\{a\})}{\sigma_i}\right]^2,
\label{eq:chi2}
\end{eqnarray}
where $N$ labels the data sets, $D_i$ denotes an individual data value with $\sigma_i$ point-to-point uncertainty (statistical and systematic uncertainties added in quadrature), and $T_i$ is the corresponding theory prediction with a parameter set $\{a\}$. If an overall (relative) normalization uncertainty $\sigma_N^{\rm norm}$ is given, the normalization factor $f_N \in [1-\sigma_N^{\rm norm},1+\sigma_N^{\rm norm}] $ is introduced, and determined for each parameter set $\{a\}$ by minimizing $\chi^2_N$. Thus, $f_N$ is computed in every round of iteration, when searching for the best overall $\chi^2$. The final $f_N$ is thus an output of the analysis.

The overall normalization uncertainty is the simplest example of correlated uncertainties in the data. If the experiment is able to provide a correlation matrix accounting for more complex correlations, the definition of the $\chi^2$ should be extended further, as done e.g. in \cite{Stump:2001gu}. However, the definition (\ref{eq:chi2}) is adequate for our purposes, since such correlation matrices are presently not available for the nuclear data we use.

The weight factors $w_N$ (with default value $w_N=1$) are needed to amplify constraints provided by those data sets whose contribution to $\chi^2$ would otherwize be too small to be noticed by {\ttfamily MINUIT}. As the weights apparently induce a piece of subjectivity to the analysis, some further explanation here is in order: Weighting is justified if there is no significant mutual tension between the weighted data set and the others. In this case the weights do not introduce a strong bias in the analysis (such bias was found e.g. in \cite{Eskola:2008ca}). For example, the $\pi^0$-data makes only about 1\% of the total $\chi^2$, and changes in such a small fraction are practically buried in the numerical noise due to statistical fluctuations in the wealth of other data. Thus, inclusion of these data with a default weight of one would be effectively the same as leaving it completely out from the analysis, which would mean loosing all the large-$x$ gluon constraints it can provide. Consequently, the gluon EMC-parameter $y_e$ would need to be fixed by hand which, we think, would induce clearly more subjectivity to the fit than assigning these data some extra weight and keeping $y_e$ free. Similarly, weighting the DY data enhances its residual sensitivity to the large-$x$ sea quarks, which enables us to free the large-$x$ sea parameter $y_e$ in a controlled manner. We see these as clear improvements over the previous nPDF analyses.

In practice, however, finding a reasonable set of weights is largely an iterative procedure\footnote{For a weight determination in a neural network approach, see \cite{DelDebbio:2007ee}.}: having first found a converging fit with a limited amount of free parameters and no additional weights, it is often easy to notice whether there are features in the data sets not satisfactorily reproduced. These data sets are then assigned some extra weight and, if needed, some previously freezed parameter can be freed and the minimization procedure repeated. In this way we have arrived at the weights listed in Table~\ref{Table:Data}.

The gluons at small $x$ are only weakly constrained by the available data, and we have observed that some parameters easily drift --- especially in the error analysis --- to an unphysical region where shadowing would get stronger towards smaller nuclei. We cure this unwanted feature by adding a further term
\begin{equation}
 1000 \left[\left(y_0^G({\rm He})-y_0^G({\rm Pb})\right) - \left(y_0^S({\rm He})-y_0^S({\rm Pb})\right)\right]^2
\end{equation}
to the $\chi^2$, which effectively drives the $A$-dependence, but not the magnitude, of $R_G$ at the smallest-$x$ similar to what there is in the sea quarks. We stress that inclusion of this term to $\chi^2$ does not affect the agreement with the data in any way, but it secures a systematic deepening of the gluon shadowing when going from lighter nucleus to a heavier one, especially so for the error-sets discussed below.

\subsection{nPDF Errors}
\label{sec:nPDFUncertainties}

As mentioned earlier, one of the main goals of the present work is to develop a general-use tool for estimating the propagation of the nPDF uncertainties to any hard process quantity $X$. This brings the uncertainty analysis of nPDFs up to a similar level as in the free proton case.

In this paper we will only consider uncertainties originating from the experimental errors in the fitted data. Although the set of parameters $\{a^0\}$ giving the minimum $\chi^2$ represents our best estimate for the nPDFs, the presence of experimental uncertainties allows us to move in the neighborhood of $\{a^0\}$ without immediately deteriorating the agreement with individual data sets. To find this relevant neighborhood and to quantify the relative differences to our central nPDFs properly, we adopt the Hessian method\footnote{For alternative methods see e.g. Refs. \cite{Stump:2001gu,Ball:2008by}.}. The details of this this method are documented in Appendix~\ref{sec:ErrorAnalysis}, we summarize here only the main points.

The Hessian method rests on a quadratic approximation for a departure of $\chi^2$ from its minimum $\chi^2_0$
\begin{equation}
\chi^2 - \chi^2_0 \approx \sum_{ij} \frac{1}{2} \frac{\partial^2 \chi^2}{\partial a_i \partial a_j} (a_i-a_i^0)(a_j-a_j^0) \equiv \sum_{ij} H_{ij}(a_i-a_i^0)(a_j-a_j^0),
\label{eq:chi_2approx}
\end{equation}
which defines the Hessian matrix $H$. In constructing the expansion (\ref{eq:chi_2approx}), special attention should be paid to multi-parameter correlations. If several parameters control the same parton type in a certain $x$-region, there are usually notable correlations between them. This may lead to over-estimating the uncertainty in that region when applying the approximation in Eq.~(\ref{eq:chi_2approx}). In particular, changes in the valence quark modification $R_V$ obtained by small variations of the Fermi-motion parameter $\beta$ can be largely realized by varying $x_e$ and $y_e$ instead. In this sense $\beta$ is redundant in the error analysis and we eliminate it from the expansion (\ref{eq:chi_2approx}). The same argument applies to $A$-dependence of these parameters. The 15 final parameters considered in our error analysis are indicated in Table~\ref{Table:Params}.

A practical and numerically more stable way is to work not with the original parameters $\{a\}$, but use a basis $\{z\}$ which diagonalizes the Hessian matrix. Assuming linear propagation of uncertainties, the symmetric extreme deviation in any quantity $X$ that corresponds to a given rise $\Delta \chi^2$ in $\chi^2$, can be written as
\begin{equation}
(\Delta X)^2 \approx \frac{1}{4} \sum_k \left( X(S^+_k)-X(S^-_k) \right)^2,
\label{eq:SymmetricPDFerrors}
\end{equation}
where $X(S^+_k)$ and $X(S^-_k)$ denote the values of the quantity $X$ computed by the nPDF sets $S_k^+$ and $S_k^-$, which are obtained by displacing the fit parameters in the $z$-space direction $k$ such that $\chi^2$ grows by a certain fixed amount $\Delta \chi^2$. Lower and upper uncertainties $\Delta X^\pm$ can be computed also separately, using a prescription \cite{Nadolsky:2001yg}
\begin{eqnarray}
(\Delta X^+)^2 & \approx & \sum_k \left[ \max\left\{ X(S^+_k)-X(S^0), X(S^-_k)-X(S^0),0 \right\} \right]^2 \label{eq:ASymmetricPDFerrors} \\
(\Delta X^-)^2 & \approx & \sum_k \left[ \max\left\{ X(S^0)-X(S^+_k), X(S^0)-X(S^-_k),0 \right\} \right]^2 \nonumber,
\end{eqnarray}
where $S^0$ denotes the central PDF set giving the minimum $\chi^2$. All the errors shown in this paper have been computed by the latter method, with $\Delta \chi^2 = 50$  which corresponds to a ``90\% confidence criterion'' defined  in Appendix~\ref{sec:ErrorAnalysis}. As discussed in Sec.~\ref{sec:results} the Hessian matrix is computed for 15 fitting parameters in our analysis. According to the procedure explained above, this leads to 30 error-sets in addition to the best-fit central set. All these 31 nPDF sets needed for the evaluation of Eq.~(\ref{eq:ASymmetricPDFerrors}) are included in our \texttt{EPS09} release \cite{EPS09code}. This allows the computation of uncertainties in nuclear cross-section ratios, similar to those in Eq.~(\ref{RF2RDY}), by any interested user. The corresponding calculation of uncertainty estimates for the absolute cross-sections (instead of ratios), will be explained later in Sec.~\ref{sec:Application}.

In this paper we do not discuss the possible shortcomings concerning our whole framework such as breakdown of pQCD  factorization at small $Q^2$ or small/large-$x$ corrections to the DGLAP splitting functions already at the nucleon-nucleon level collisions --- for an extensive study of such issues, we recommend the reader to consult Refs. \cite{Martin:2003sk,Dittmar:2009ii}. Also, the presence of additional $(Q^2)^{-n}$ power corrections are expected to become increasingly important for large nuclei, see e.g. \cite{Qiu:2003vd,Gribov:1984tu,Mueller:1985wy,Kang:2007nz,Kulagin:2004ie}. As we will see next, however, the present framework seems to work very well in the kinematical domain considered in this analysis.

\section{Results and Discussion}
\label{sec:results}

\subsection{Final parametrization}
\label{subsec:finalparameters}

\begin{table}[!htb]
\begin{tabular}{p{1cm}|c|lll}
 {\bf NLO}  & Param.   	&  Valence  $R_V^A$    	&  Sea $R_S^A$          	&  Gluon $R_G^A$\\
\hline
\hline
 1 &  $y_0$	   & {Baryon sum rule}	  &  \textbf{0.785}       	
&  {Momentum sum rule}	\\
 2 &  $p_{y_0}$ & 	---		 	  &  \textbf{-0.136}  	&   ---	\\ 
 3 &  $x_a$    	   &  {\textbf{6.56 $\times {\bf 10^{-2}}$}} &  {\textbf{8.74 $\times {\bf 10^{-2}}$}}        &  0.1 {fixed} 
\\
 4 &  $p_{x_a}$    &  0, {fixed}  &  0, {fixed}  &  0, {fixed}   \\
 5 &  $x_e$    	   &  \textbf{0.688}         & {as valence}  &  {as valence} \\
 6 &  $p_{x_e}$    &  0, {fixed}    &  0, {fixed}   &  0, {fixed}   \\
 7 &  $y_a$    	   &  \textbf{1.05}        	&  \textbf{0.970}       	&  \textbf{1.207}     \\
 8 &  $p_{y_a}$    &  \textbf{1.47 $\times {\bf 10^{-2}}$}   	& \textbf{-8.12 $\times {\bf 10^{-3}}$} & \textbf{2.72 $\times {\bf 10^{-2}}$} \\
 9 &  $y_e$   &  \textbf{0.901}  &  \textbf{1.076}  & \textbf{0.625} \\
 10 &  $p_{y_e}$    & \textbf{-2.81 $\times {\bf 10^{-2}}$}    &  {as valence}  &  {as valence} \\
 11 &  $\beta$  	   & {1.31} 	& 1.3, {fixed}     &  1.3, {fixed} \\
 12 &  $p_{\beta}$  &  {4.63 $\times {10^{-2}}$}   & 0, {fixed}     &  0, {fixed}	\\
\hline
\hline
 {\bf LO}  &    	&      	&            	&   \\
\hline
 1 &  $y_0$	   & {Baryon sum rule}	  &  \textbf{0.890} &  {Momentum sum rule}	\\
 2 &  $p_{y_0}$ & 	---		 	  &  \textbf{-8.03 $\times {\bf 10^{-2}}$}	&   ---	\\ 
 3 &  $x_a$    	   &  {\textbf{6.84 $\times {\bf 10^{-2}}$}} &  {\textbf{0.127}} &  0.1 {fixed} \\
 4 &  $p_{x_a}$    &  0, {fixed}  &  0, {fixed}  &  0, {fixed}   \\
 5 &  $x_e$    	   &  \textbf{0.727}         & {as valence}  &  {as valence} \\
 6 &  $p_{x_e}$    &  0, {fixed}    &  0, {fixed}   &  0, {fixed}   \\
 7 &  $y_a$    	   &  \textbf{1.04}        	&  \textbf{0.971}       	&  \textbf{1.096}     \\
 8 &  $p_{y_a}$    &  \textbf{1.53 $\times {\bf 10^{-2}}$}   	& \textbf{-1.65 $\times {\bf 10^{-2}}$} & \textbf{4.57 $\times {\bf 10^{-2}}$} \\
 9 &  $y_e$   &  \textbf{0.902}  &  \textbf{1.062}  & \textbf{0.901} \\
10 &  $p_{y_e}$    & \textbf{-2.97 $\times {\bf 10^{-2}}$}    &  {as valence}  &  {as valence} \\
11 &  $\beta$  	   & {1.34} 	& 1.3, {fixed}     &  1.3, {fixed} \\
12 &  $p_{\beta}$  & {5.07 $\times {10^{-2}}$}   & 0, {fixed} &  0, {fixed} \\
\hline
\hline
\end{tabular}
\caption[]{\small List of parameters defining the modifications $R_V^A$, $R_S^A$ and $R_G^A$ through Eqs.~(\ref{eq:parametrization}) and (\ref{eq:Adependence}) at our initial scale $Q_0^2=1.69$~GeV$^2$ in NLO and LO. The final values for the 15 free parameters considered in the minimization procedure and in our error analysis are shown in bold face. The values for the parameters $\beta$ and $p_{\beta}$ for valence quarks were fixed on the basis of a 17 parameter fit.}
\label{Table:Params}
\end{table}
\begin{figure}[!htb]
\center
\includegraphics[scale=0.5]{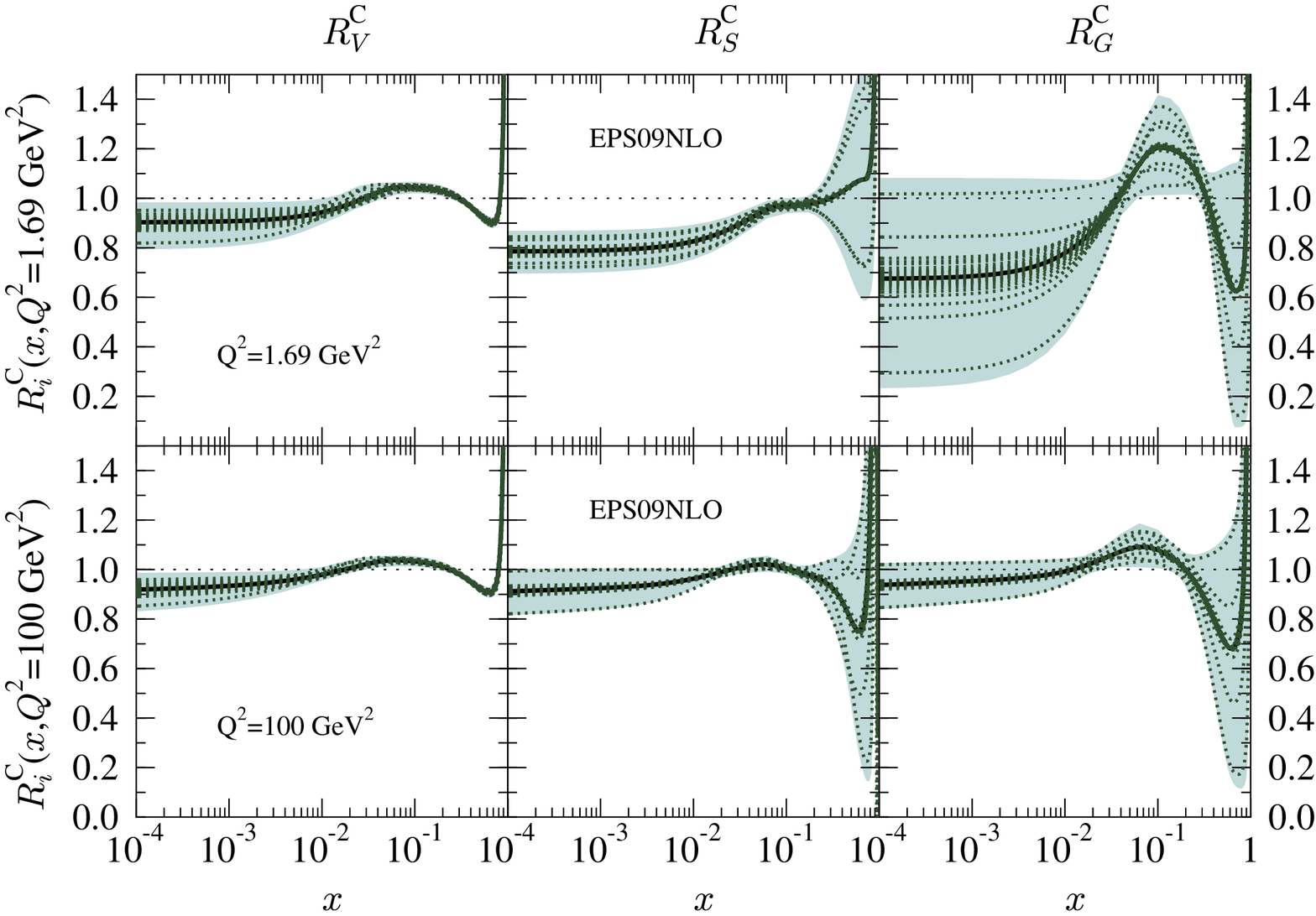} \\
\vspace{-0.5cm}
\includegraphics[scale=0.5]{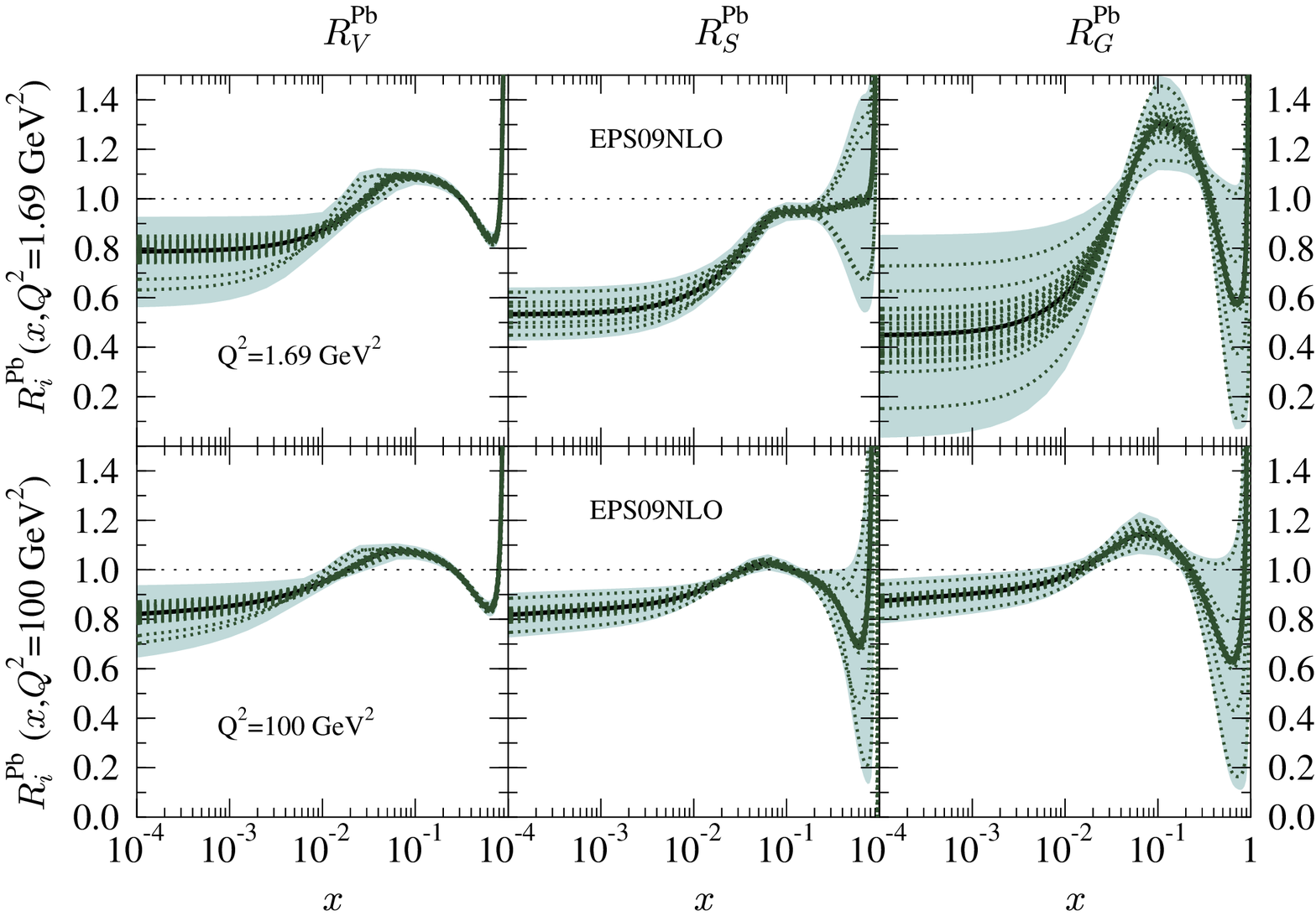}
\caption[]{\small The nuclear modifications $R_V$, $R_S$, $R_G$ for Carbon (upper group of panels) and Lead (lower group of panels) at our initial scale $Q^2_0=1.69 \, {\rm GeV}^2$ and at $Q^2=100 \, {\rm GeV}^2$. The thick black lines indicate the best-fit results, whereas the dotted green curves denote the error sets. The shaded bands are computed from Eq.~(\ref{eq:ASymmetricPDFerrors}).}
\label{Fig:AllPDFs}
\end{figure}

In Table \ref{Table:Params} we present the values of the 17 parameters resulting from the $\chi^2$ minimization and those that we fixed. The error analysis was performed for the 15 parameters shown in bold face. The corresponding nuclear modifications together with the uncertainty estimates at our initial scale $Q^2_0=1.69 \, {\rm GeV}^2$ and at a higher scale $Q^2=100 \, {\rm GeV}^2$ are shown for Carbon and Lead in Fig.~\ref{Fig:AllPDFs}. We make the following observations for gluons, sea quarks and valence quarks respectively.

At our parametrization scale $Q^2_0$ there are large uncertainties in both small-$x$ and large-$x$ gluons. Only at moderate $x$ the gluons are somewhat better controlled as the precision small-$x$ DIS data --- although directly more sensitive to the sea quarks --- constrain the gluons at slightly higher $x$ due to the parton branching encoded into DGLAP evolution. At higher $Q^2$ the small-$x$ uncertainty rapidly shrinks whereas at large $x$ a sizable uncertainty band persists. 

The narrow throat in the sea quark uncertainty band at $x \sim 10^{-2} \ldots 10^{-1}$ reflects the good constraining power of the precision DIS data. Towards higher $x$, the uncertainty grows as the accuracy of the DY data is not enough to decisively nail down nuclear modification for the sea quarks there. Note, however, that unlike in our earlier works, the parameter $y_e$ was free. Towards small $x$ the errors are perhaps surprisingly small given that there are no direct data constraints. This is an artefact of the chosen form of the fit function as the tight constraints at $x \sim 10^{-2} \ldots 10^{-1}$ fix also the smaller-$x$ behaviour leading to an unrealistically small uncertainty band there. This should be kept in mind when computing hard cross-sections involving very-small-$x$ gluons.

The valence quark modifications are well under control at $x \gtrsim 0.1$ --- thanks to the DIS data probing the valence quarks at this region. At small-$x$, the baryon number sum rule together with the chosen form of the fit function halts the uncertainty band from growing, although there are no data that would directly constrain the valence quarks there.

\subsection{Comparison with the data}
\label{subsec:compdata}

\begin{figure}[htbp]
\center
\subfigure{
\includegraphics[scale=0.39]{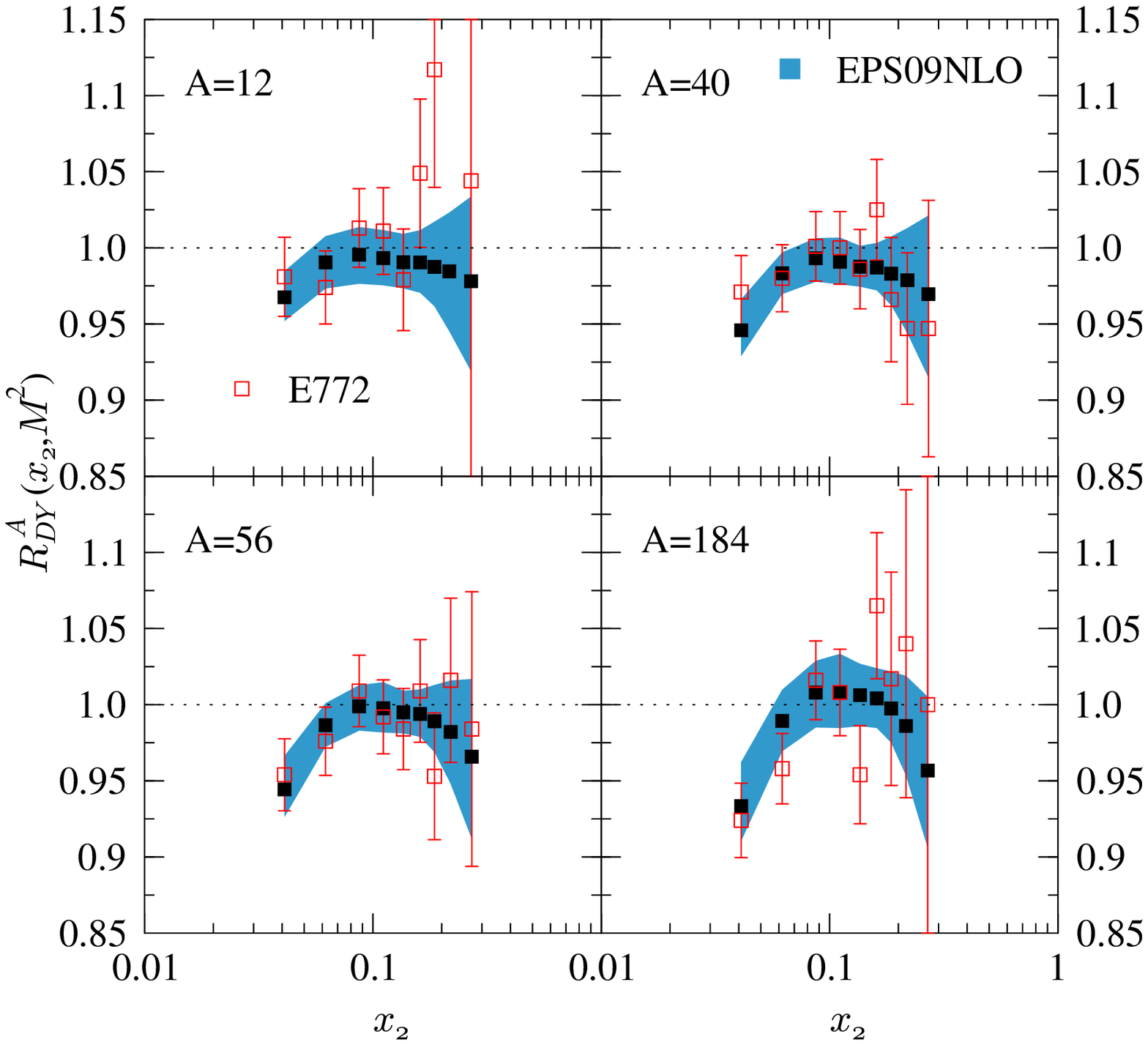}} \hspace{-1.0cm}
\subfigure{
\includegraphics[scale=0.40]{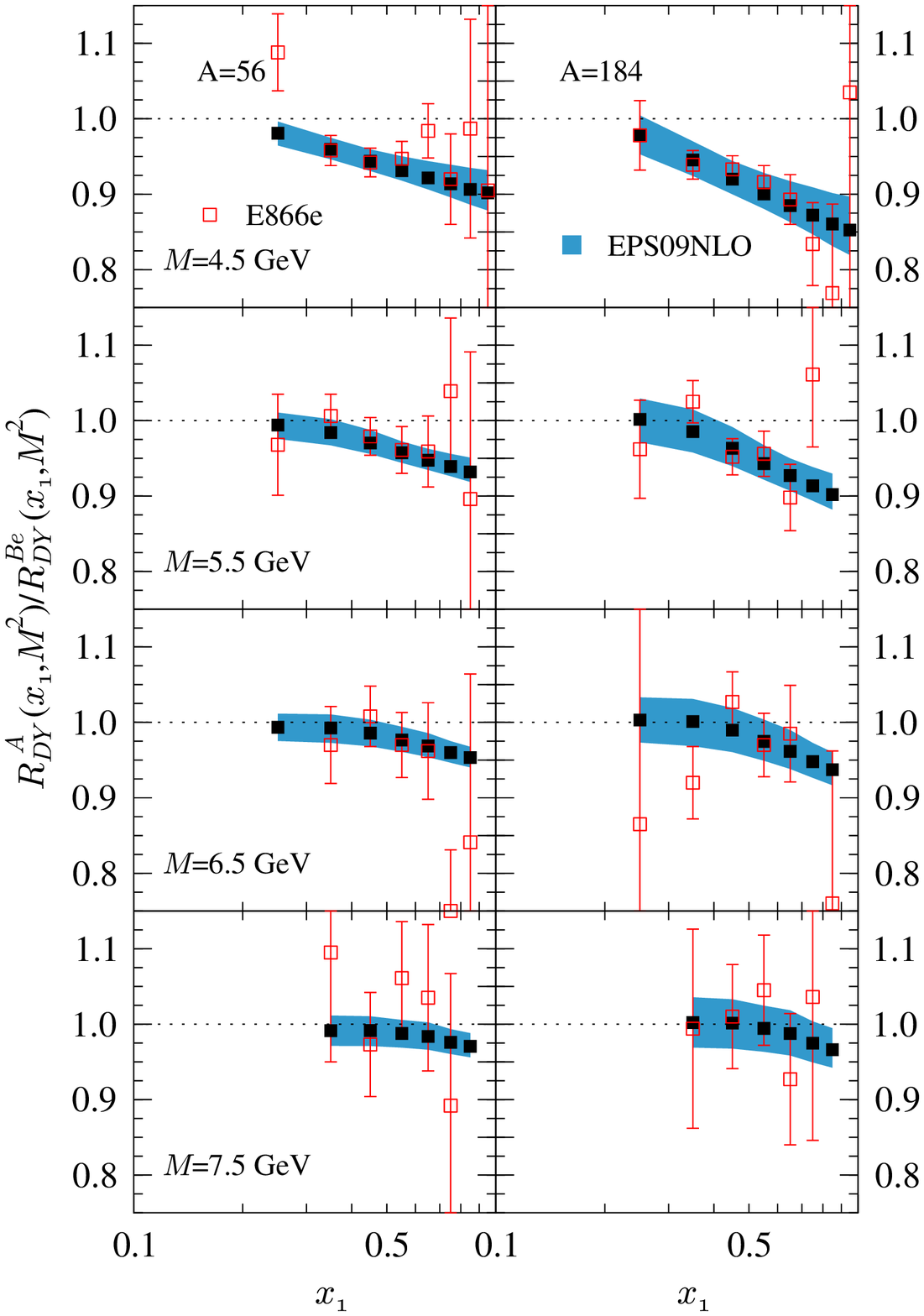}}
\caption[]{\small The computed NLO $R_{\rm DY}^{\rm A}(x,M^2)$ (filled squares and error bands) as a function of $x_2$ (left) and $x_1$ (right) compared with the E772 \cite{Alde:1990im} and E866 \cite{Vasilev:1999fa} data (open squares).}
\label{Fig:RDY}
\end{figure}
\begin{figure}[!htbp]
\center
\subfigure{
\includegraphics[scale=0.45]{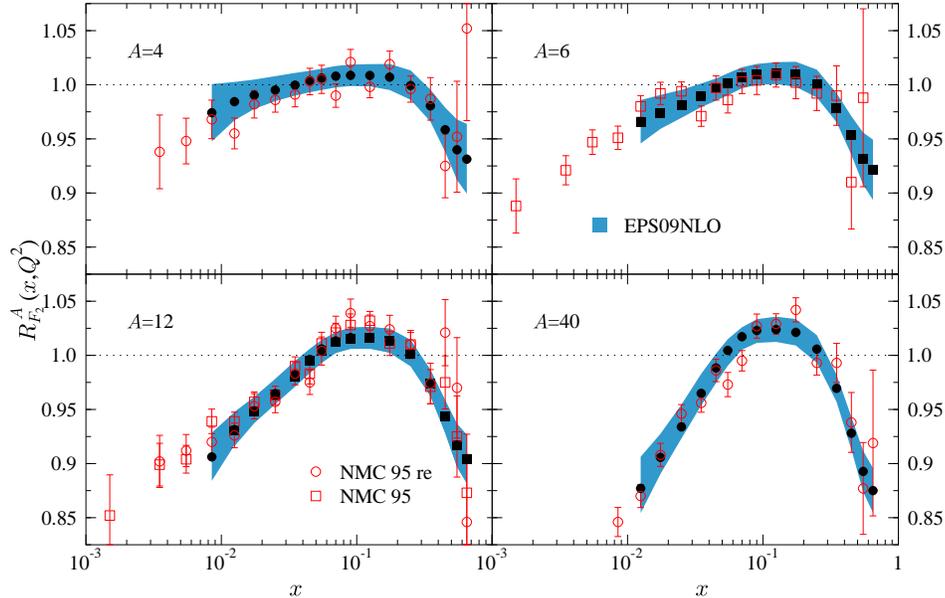}}
\caption[]{\small The calculated NLO $R_{F_2}^A(x,Q^2)$ and (filled symbols and error bands) compared with the NMC 95 \cite{Arneodo:1995cs} and the reanalysed NMC 95 \cite{Amaudruz:1995tq} data (open symbols).}
\label{Fig:RF2A1}
\end{figure}
\begin{figure}[!htb]
\center
\subfigure{
\includegraphics[scale=0.35]{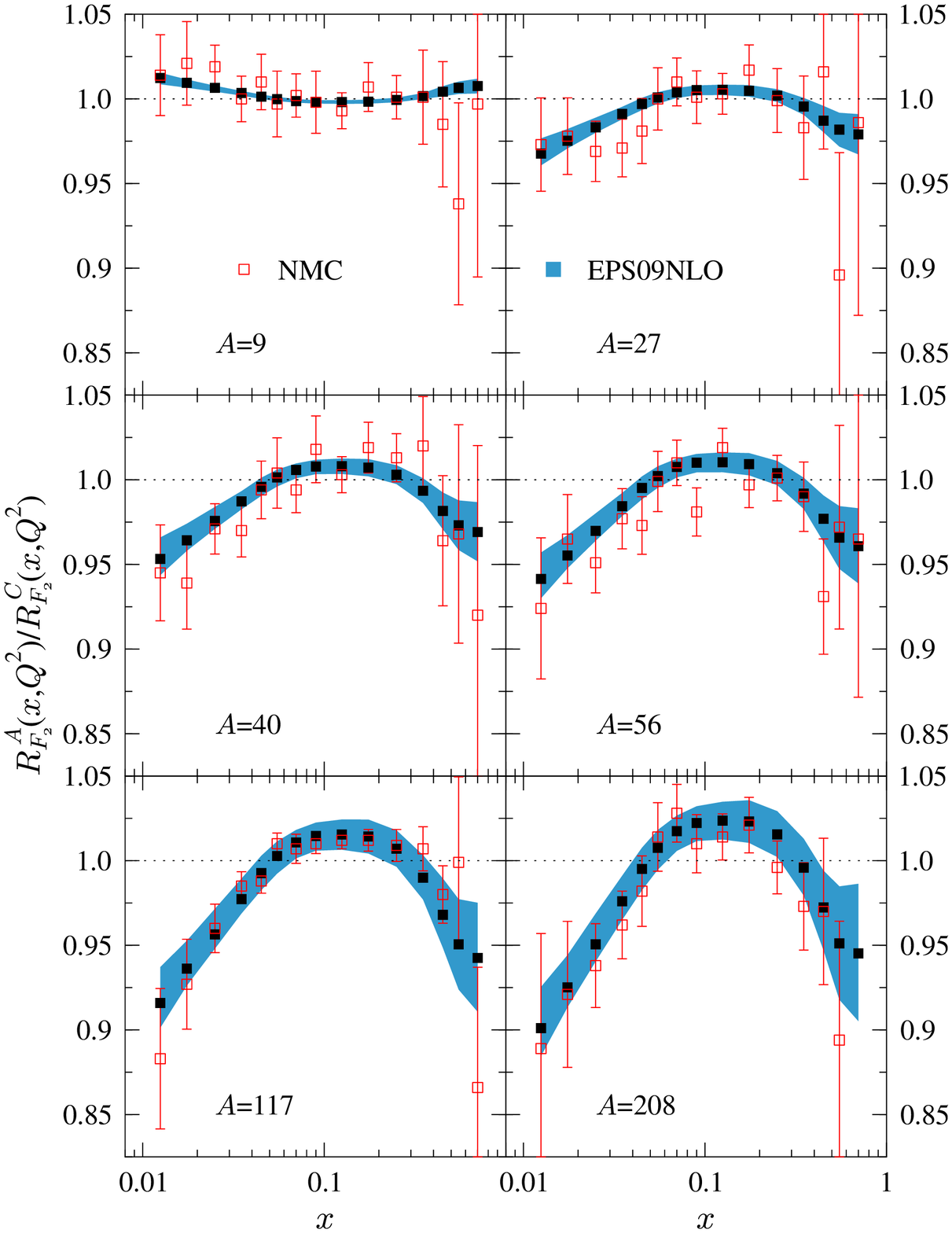}}
\subfigure{
\includegraphics[scale=0.35]{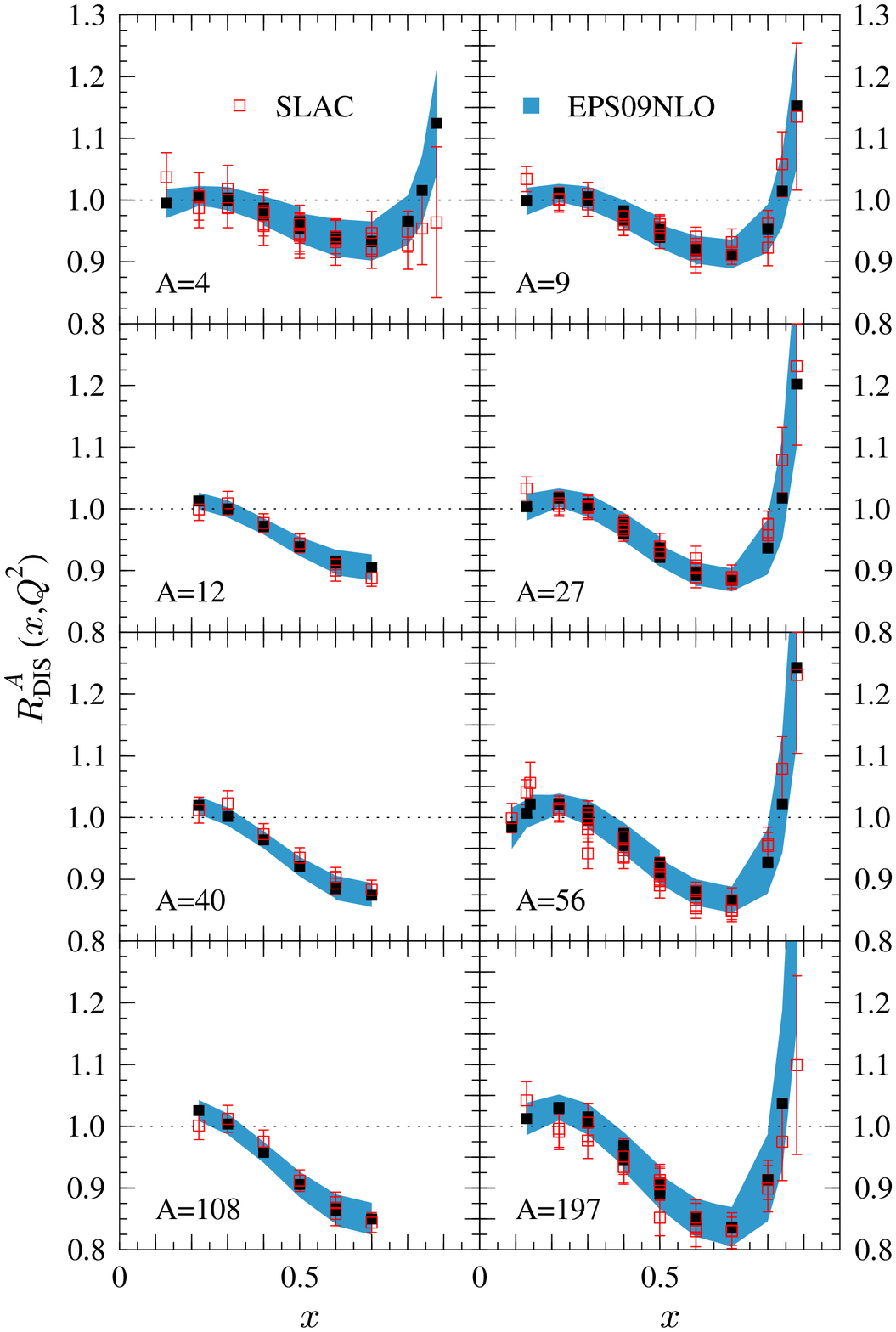}}
\caption[]{\small The computed NLO ratios $R_{F_2}^A(x,Q^2)$ vs. $R_{F_2}^{\mathrm C}(x,Q^2)$ and $R_{\rm DIS}^A(x,Q^2)$ (filled squares and error bands) compared with the NMC \cite{Arneodo:1996rv} (left) and SLAC \cite{Gomez:1993ri} (right) data (open squares).}
\label{Fig:RF2A2}
\end{figure}

In this section we demonstrate how the obtained NLO parametrization reproduces the data and how the nPDF uncertainties in Fig.~\ref{Fig:AllPDFs} relate to the error bars of the data. In all figures we present, the open symbols with error bars denote the experimental data and the filled symbols or thicker lines are the corresponding calculated values. The shaded light blue bands indicate the uncertainty estimates computed using our 30 error sets in Eq.~(\ref{eq:ASymmetricPDFerrors}). Unless otherwize stated, the point-to-point systematic and statistical errors have been added in quadrature.

 In Fig.~\ref{Fig:RDY} we show the Drell-Yan data. As mentioned earlier, we have found that there is some dependence on the large-$x$ sea quarks in the E772 data shown in the left panels. However, the apparent large scatter from one nucleus to another makes it difficult to well exploit this sensitivity and completely impossible to extract the $A$-dependence of the large-$x$ $R_S$ independently. The growing uncertainty band towards larger $x$ is nicely in line with the growing large-$x$ uncertainty for $R_S$ in Fig.~\ref{Fig:AllPDFs}. We would like to emphasize two systematic features present in the data, which are well caught by our global analysis: The downward trend (shadowing) of the smallest-$x$ points in the E772 data, and the diminishing nuclear effects towards larger invariant mass $M$ in the E886 data.

\begin{figure}[!htb]
\center
\includegraphics[scale=0.5]{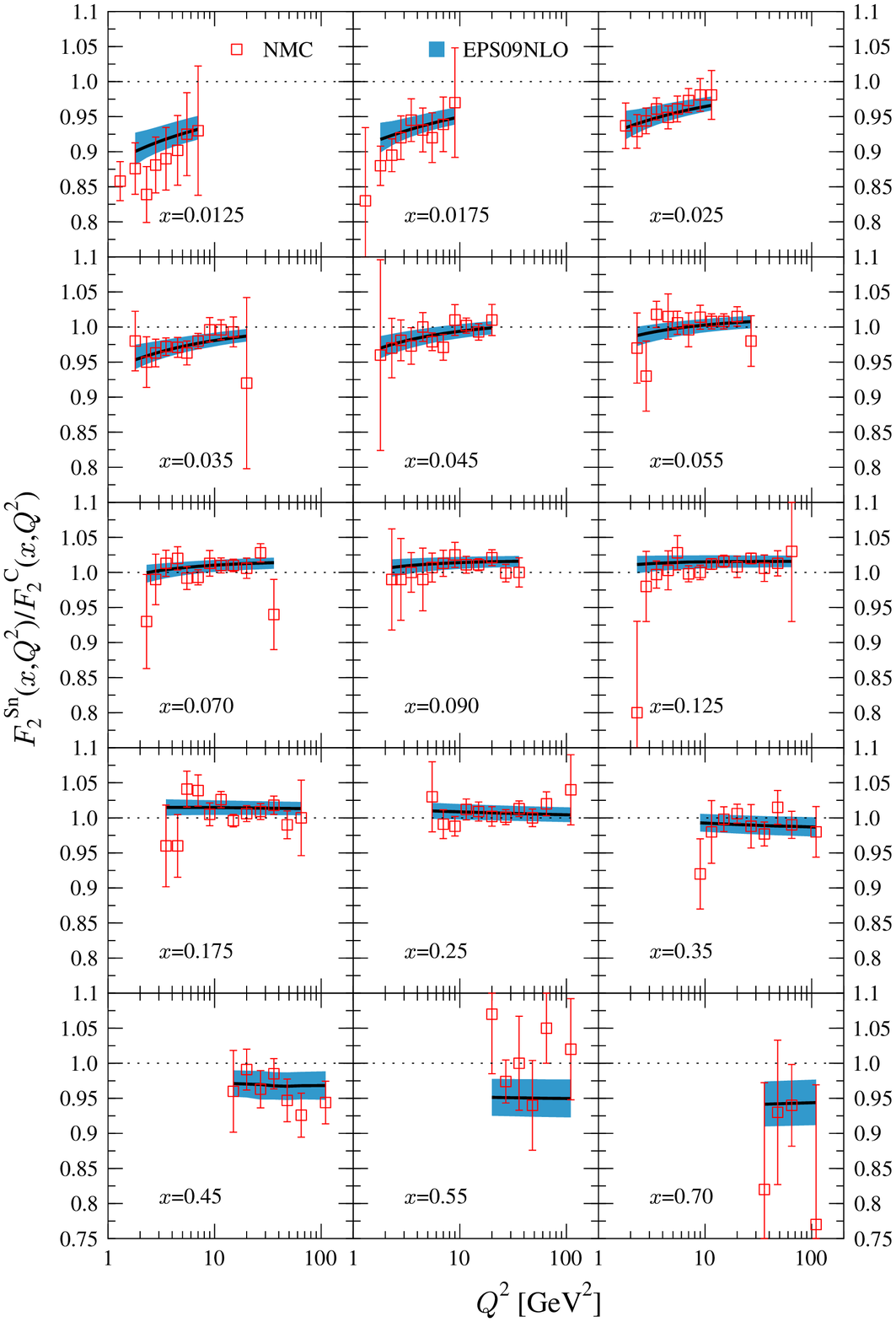}
\caption[]{\small The calculated NLO scale evolution (solid black lines and error bands) of the ratio
  $F_2^{\mathrm{Sn}}/F_2^{\mathrm{C}}$ compared with the NMC data
  \cite{Arneodo:1996ru} for  several fixed values of $x$.}
\label{Fig:RF2SnC}
\end{figure}
\begin{figure}[!htb]
\center
\subfigure{
\includegraphics[scale=0.35]{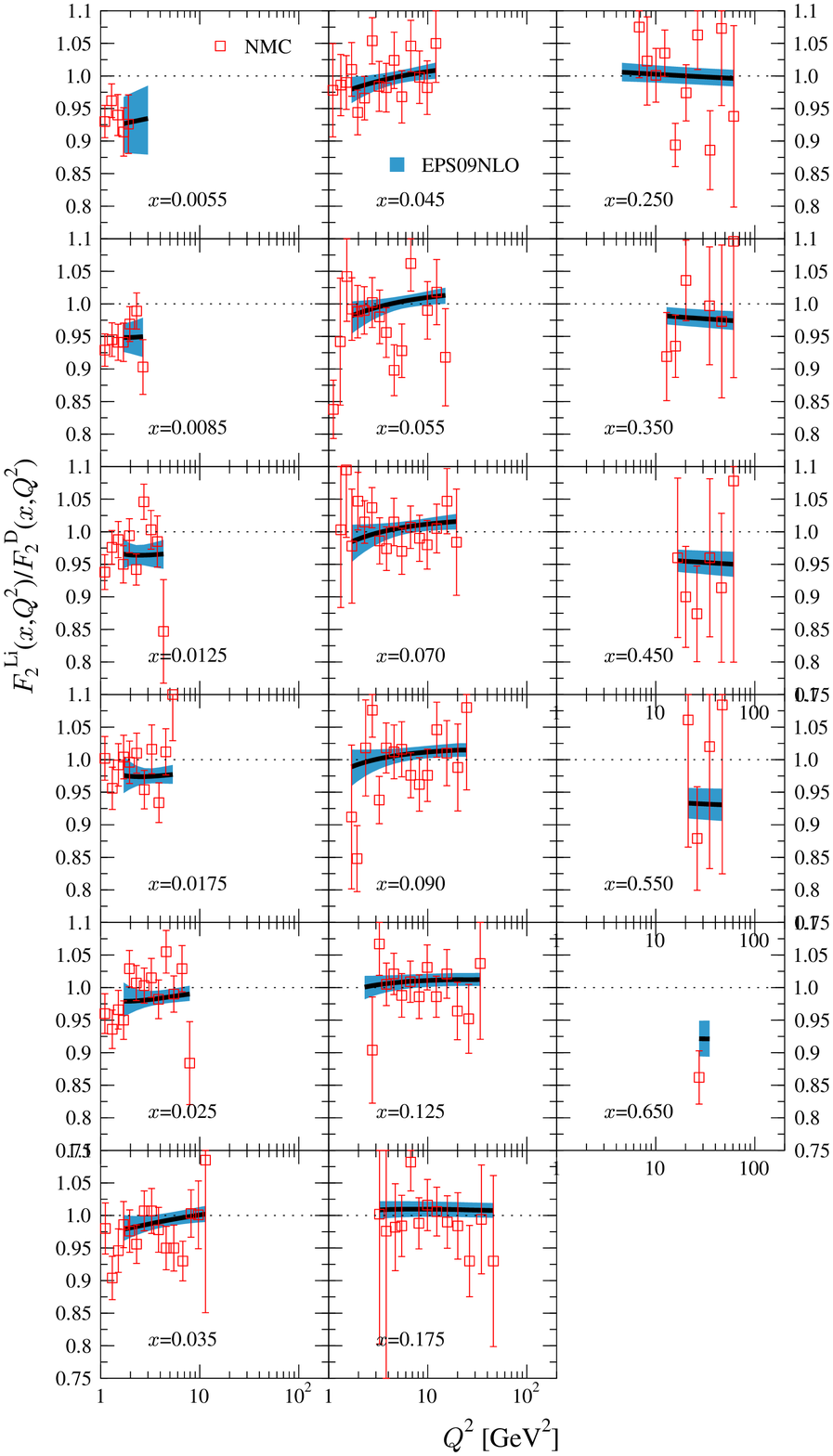}}
\subfigure{
\includegraphics[scale=0.35]{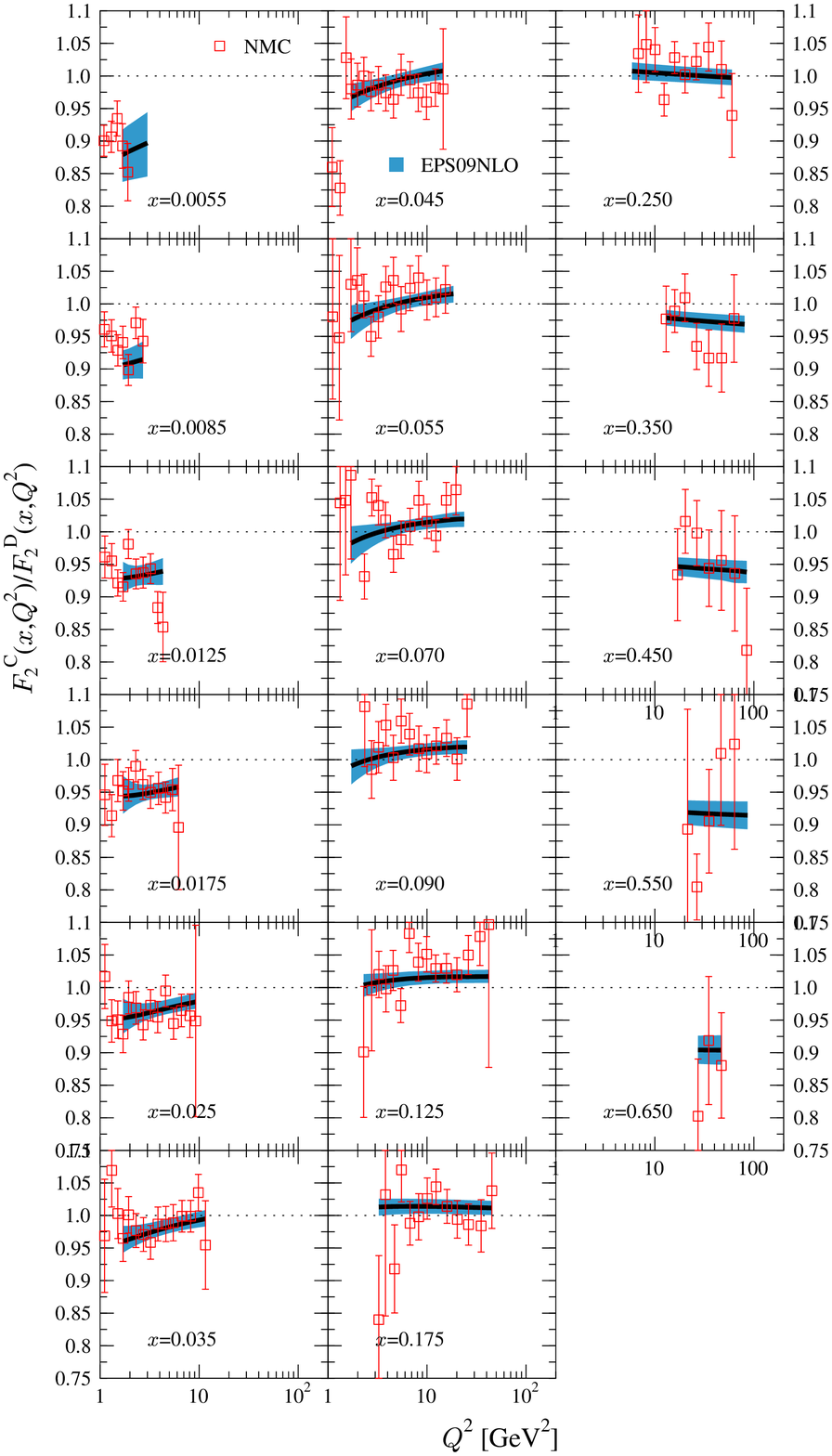}}
\caption[]{\small The calculated NLO scale evolution (solid black lines and error bands) of the ratios $F_2^{\mathrm{Li}}/F_2^{\mathrm{D}}$ and $F_2^{\mathrm{C}}/F_2^{\mathrm{D}}$ compared with the NMC data \cite{Arneodo:1996ru} for  several fixed values of $x$. One percent systematic error has been added in quadrature with the statistical errors.}
\label{Fig:RF2_slopes_in_Li_and_C}
\end{figure}
\begin{figure}[!h]
\centering
\includegraphics[width=18pc]{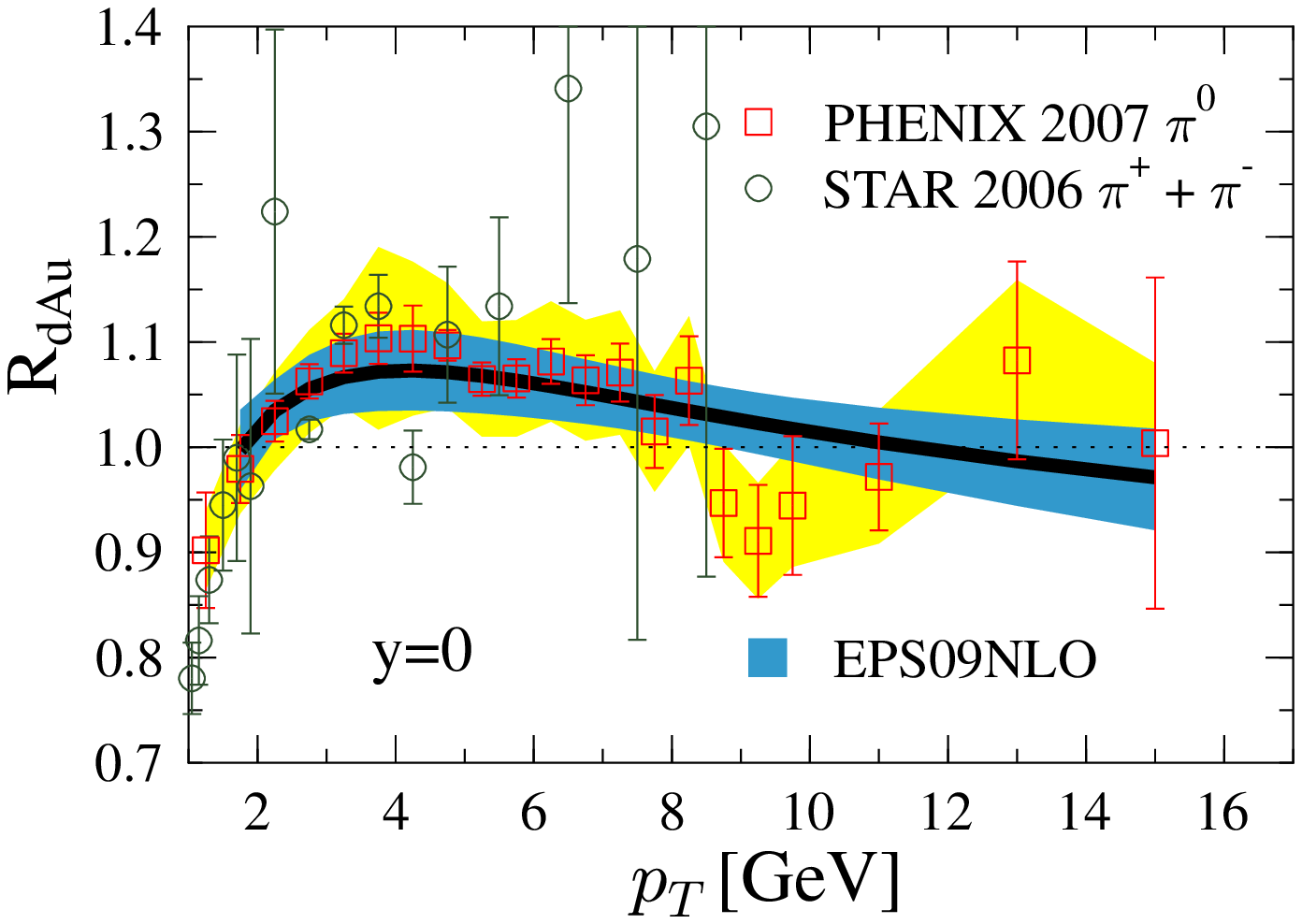}
\caption[]{\small 
{The computed $R_{\rm dAu}$ (thick black line and blue error band) at $y=0$ for inclusive pion production compared with the PHENIX \cite{Adler:2006wg} data (open squares). The error bars are the statistical uncertainties, and the yellow band indicate the point-to-point systematic errors. The additional $10\%$ overall normalization uncertainty in the data is not shown. The data have been multiplied by the optimized normalization factor $f_N = 1.03$, which is an output of our analysis. Also the STAR data \cite{Adams:2006nd} (open circles) multiplied by a normalization factor $f_N = 0.90$ are shown for comparison.}}
\label{Fig:PHENIX}
\end{figure}

The DIS data in Figs.~\ref{Fig:RF2A1}-\ref{Fig:RF2_slopes_in_Li_and_C} comprise the bulk of our experimental constraints. Since a large part of the DIS data points lies at relatively low $Q^2$ (see Fig.~\ref{Fig:KinematicReach}), the good agreement between the data and the theory also verifies that our fit function is flexible enough. At higher $Q^2$, the evolution effects become sizable and the fit function dependence is no longer as critical.

The $A$-dependence of the nuclear effects is encoded in the parametrization through Eq.~(\ref{eq:Adependence}), and its determination is mainly possible due to the existence of DIS data for several nuclei. This dependence and its correct reproduction is obvious e.g. from Figs.~\ref{Fig:RF2A1} and \ref{Fig:RF2A2} in which data from several nuclei are gathered together for an easy comparison.

In addition to the DY data, the scaling violation effects are clearly visible in the NMC data Figs.~\ref{Fig:RF2SnC} and \ref{Fig:RF2_slopes_in_Li_and_C}, which display the $Q^2$-evolution of the DIS structure function $F_2$ ratios. Thus, the DGLAP evolution plays a significant role in a successful description of these data.

In Fig.~\ref{Fig:PHENIX} we finally show the nuclear modification for inclusive pion yield in d+Au collisions at midrapidity. We plot both PHENIX $\pi^0$ \cite{Adler:2006wg} and STAR $\pi^+ + \pi^-$ \cite{Adams:2006nd} data\footnote{The STAR data was, however, not included in the fit due to poorer statistics and more restricted kinematical reach.}.  We are, at the moment, the only group that includes this type of data in the global DGLAP analysis of nPDFs. We observe, however, that the shape of the pionic $R_{\rm dAu}$ as a function of $p_T$ is very well reproduced without any additional effects, indicating that the universality of the nPDFs works well. We thus find no reason to exclude these data from the global fit analyses.

Typically, the error bands in most of the figures are roughly of the same size as (especially never badly overshooting) the error bars in the data, verifying that we have successfully propagated them to our nPDFs. In some cases, however, the size of the nPDF errors is apparently below the experimental uncertainties. This can happen for two reasons: First, in cases like Fig.~\ref{Fig:RF2_slopes_in_Li_and_C}, the small error bands result simply because there are other, more precise data sets probing the same $x$-region and, by construction, the size of our error bands are controlled by the most stringent ones. Second, in cases like the panels of Fig.~\ref{Fig:RF2A2} which compare Beryllium or Aluminium to Carbon, where cross-section ratio is taken between two nuclei very close in $A$, the errors in other parameters than those from the powers $p_{d_i}$ in Eq.~(\ref{eq:Adependence}) tend to cancel.

\subsection{Comparison with earlier global analyses}
\label{sec:Comparison_to_earlier_analyses}

\begin{figure}[!htb]
\center
\includegraphics[scale=0.55]{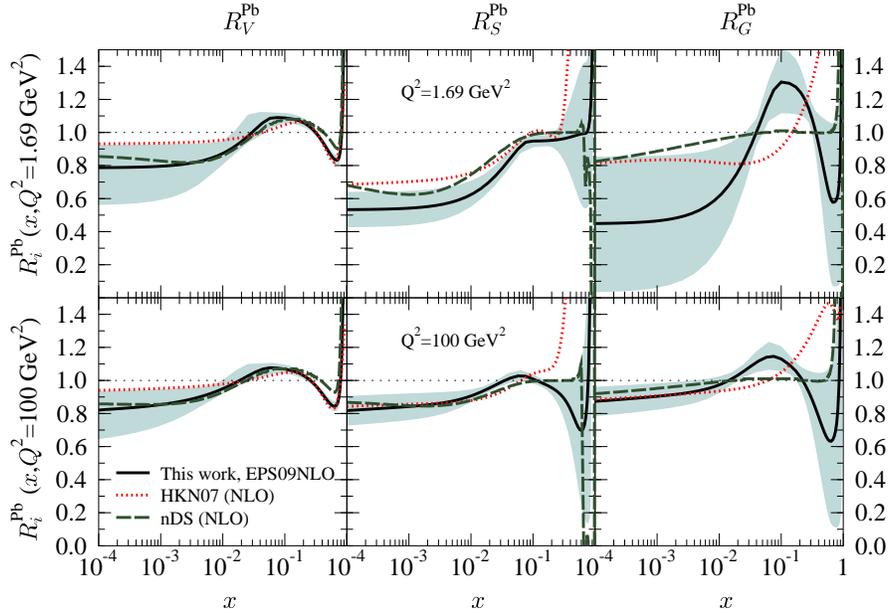}
\caption[]{\small Comparison of the average valence and sea quark, and gluon modifications at $Q^2 = 1.69 \, {\rm GeV}^2$ and $Q^2 = 100 \, {\rm GeV}^2$ for Pb nucleus from the NLO global DGLAP analyses HKN07~\cite{Hirai:2007sx}, nDS~\cite{deFlorian:2003qf} and this work, \ttfamily{EPS09NLO}.}
\label{Fig:NLOcomp}
\end{figure}

In Fig.~\ref{Fig:NLOcomp} we compare the results obtained in the present analysis \texttt{EPS09} to the earlier NLO global analyses nDS~\cite{deFlorian:2003qf} and HKN07~\cite{Hirai:2007sx}. The comparison is shown at two scales, $Q^2 = 1.69 \, {\rm GeV}^2$ and $Q^2 = 100 \, {\rm GeV}^2$, overlaid on our error bands. Evidently, the main differences in the central fits are in the gluon modifications $R_G$, especially at small $x$ and low $Q^2$, but going to higher $Q^2$ the curves tend to gather closer together. The exception is the large-$x$ region where significant differences persist. This comparison illustrates the large leftover freedom especially in nuclear gluon sector if only DIS and DY data are used as constraints.

\begin{figure}[!h]
\centering
\includegraphics[width=18pc]{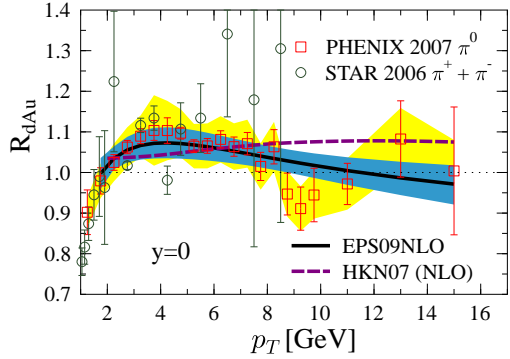}
\caption[]{\small {As Fig.~\ref{Fig:PHENIX} but also the prediction from HKN07 (NLO) is shown. The difference between EPS09 and HKN07 here demonstrates the constraining power of these data in pinning down the nuclear gluon PDFs in the mid-$x$ and large-$x$ regions.}}
\label{Fig:PHENIX_comp}
\end{figure}

The majority of the discrepancies in Fig.~\ref{Fig:NLOcomp} originate from the adopted forms of the fit functions and assumptions made about the parameters in those $x$-regions which are not well constrained by the DIS or DY data. In particular, HKN07 and nDS did not include pion d+Au data which favors the presence of an EMC suppression for gluons. In order to demonstrate the importance of these data in constraining the large-$x$ gluons, Fig.~\ref{Fig:PHENIX_comp} displays the pion $R_{\rm dAu}$ including also the prediction obtained by using the HKN07 NLO nPDFs. Although the HKN07 prediction remains largely within the experimental errors, the shape of the EPS09 curve is clearly more consistent with the data. The dominant partonic channel for inclusive pion production is this $p_T$ range the gluon-intiated one. Thus, the qualitatively contradicting behaviour between HKN07 and EPS09 in Fig.~\ref{Fig:PHENIX_comp} is inevitably linked to the differences in their nuclear effects for gluon PDFs.

\subsection{Leading-order analysis}
\label{sec:Leading order analysis}

Although the NLO analysis is the main objective in the present paper, we have also performed a new LO analysis to provide the tools for computing uncertainty estimates also in this widely-used framework. The LO framework is basically the same as in NLO, but the partonic cross-sections and DGLAP splitting functions are one power lower in $\alpha_s$, and we use the CTEQ6L1 \cite{Pumplin:2002vw} free proton PDF set as the baseline. The parameters specifying the LO fit are listed in Table~\ref{Table:Params}. 

\begin{figure}[tb]
\center
\includegraphics[scale=0.6]{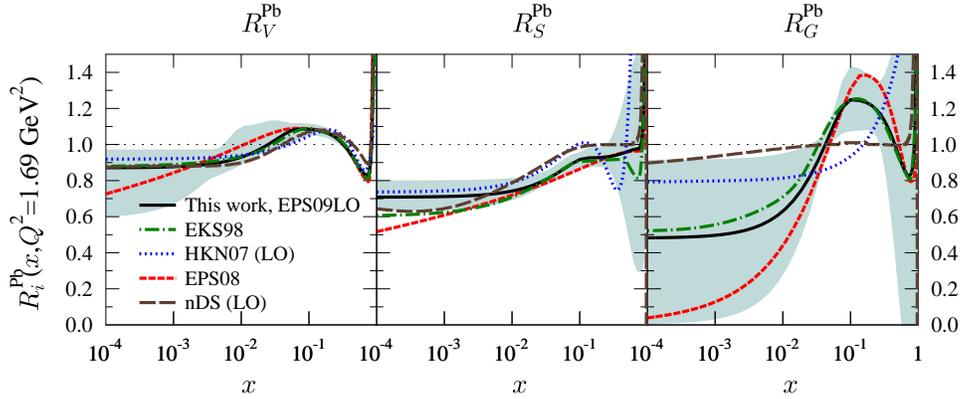}
\caption[]{\small Comparison of the average valence and sea quark, and gluon modifications at $Q^2 = 1.69 \, {\rm GeV}^2$ for Pb nucleus from LO global DGLAP analyses EKS98 \cite{Eskola:1998iy,Eskola:1998df}, EKPS \cite{Eskola:2007my}, nDS~\cite{deFlorian:2003qf}, HKN07~\cite{Hirai:2007sx}, and this work \texttt{EPS09LO}.}
\label{Fig:LOcomp}
\end{figure}

Due to the qualitatively similar behaviour of the LO and NLO fits, we do not present a detailed discussion of the LO results but only show, in Fig.~\ref{Fig:LOcomp}, the resulting nuclear modifications for Lead at the initial scale $Q_0^2$ with their $\Delta \chi^2=50$ uncertainty bands. The same figure also presents the comparison to other LO analyses. We find that our error bands can accommodate most of the earlier analyses. So, in practice, using the \texttt{EPS09LO} error sets all these different nPDF parametrizations are effectively covered.

 As can be read off from Table~\ref{Table:Data} there is practically no difference between the LO and NLO fits at the level of the values of $\chi^2$ obtained. Intuitively, however, the NLO fit should be better constrained than the LO one as there are more partonic channels open in the NLO DIS and DY cross-sections. Indeed this expectation is confirmed by Figs.~\ref{Fig:AllPDFs} and \ref{Fig:LOcomp}. Especially the large-$x$ gluons are better constrained at NLO.

\section{Application}
\label{sec:Application}

In this section we apply the obtained \texttt{EPS09NLO} parametrization --- the central set and 30 error sets --- to a cross-section that was not included in the fit. Through this example we also want to demonstrate how our parametrization should be applied in practice.

We consider here inclusive negative hadron $h^-$ production at forward (pseudo) rapidities $\eta=2.2$ and $\eta=3.2$, in p+p and d+Au collisions, measured by the BRAHMS collaboration \cite{Arsene:2004ux} at RHIC. In our previous article \cite{Eskola:2008ca} we discussed how the suppression observed in the nuclear modification $R_{\rm dAu}$ obtained from these data would strongly favour very deep gluon shadowing, and we searched for the strongest possible one that would still not contradict the available DIS and DY data. The analysis \cite{Eskola:2008ca} was performed in a LO framework and we were forced to use fragmentation functions for average hadrons $h^+ + h^-$ instead of charge-separated ones for $h^-$ only\footnote{The extraction of the charge separated fragmentation functions from p+p data is reliable only at NLO due to significant perturbative $\mathcal O \left( \alpha_s^2 \right)$ corrections.}. In the current NLO setup we relax such simplification and employ the charge-separated NLO fragmentation functions by Sassot et al. \cite{de Florian:2007hc}.

We first investigate how well the NLO pQCD calculation can reproduce the shape and magnitude of the differential $h^-$ yields measured by BRAHMS in p+p and d+Au collisions from which the nuclear modification $R_{\rm dAu}$ is computed. The inclusive yields are linked to the cross-sections by
\begin{equation}
\frac{d^2 N^{\rm pp}}{dp_T dy} \stackrel{\rm min. bias}{=} \frac{1}{\sigma_{NN}^{\rm inelastic}} \frac{d^2\sigma^{\rm pp}}{dp_T dy}
\quad ; \quad
\frac{d^2 N^{\rm dAu}}{dp_T dy} \stackrel{\rm min. bias}{=} \frac{{\langle N_{\rm coll}\rangle}}{\sigma_{NN}^{\rm inelastic}} \frac{\frac{1}{2A}d^2\sigma^{\rm dAu}}{dp_T dy},
\label{eq:BRAHMScrosssection}
\end{equation}
where $\sigma_{NN}^{\rm inelastic}$ is the total inelastic nucleon-nucleon cross-section and ${\langle N_{\rm coll}\rangle}$ is the average number of binary nucleon-nucleon collisions. Here we take $\sigma_{NN}^{\rm inelastic} = 40 \, {\rm mb}$ \cite{Amsler:2008zzb}, and ${\langle N_{\rm coll}\rangle} = 7.2$ \cite{Arsene:2004ux}, understanding that there is an inherent uncertainty in both: $\sigma_{NN}^{\rm inelastic}$ has not been measured for $\sqrt{s} = 200 \, {\rm GeV}$ and determining its value requires extrapolation. In turn, ${\langle N_{\rm coll}\rangle}$ is a model-dependent number resulting from a Glauber-model simulation performed by the experiment. However, both of these quantities are only multiplicative overall factors and do not affect the shape of the obtained distribution.

Whenever comparing theoretical predictions to experimental data, we emphasize the importance of presenting also the theoretical uncertainties deriving from the PDF errors. For absolute cross-sections in p+p collisions these are obtained from Eq.~(\ref{eq:ASymmetricPDFerrors}) using the 40 error sets provided by CTEQ \cite{Stump:2003yu}. In the d+Au case there is an additional contribution from the uncertainties of the nuclear effects which we have quantified in the present analysis by the 30 error sets with the CTEQ central set as a baseline. In this case, we compute the d+Au cross-sections separately with 1+40+30=71 parton sets defined by:
\begin{eqnarray}
R^A_{S^0} & \times & f_{\rm Central \, set}^{\rm CTEQ6.1M} \nonumber \\
R^A_{S^0} & \times & f_{\rm Error \, set \, \pm 1}^{\rm CTEQ6.1M} \nonumber \\
R^A_{S^0} & \times & f_{\rm Error \, set \, \pm 2}^{\rm CTEQ6.1M} \nonumber \\
& \vdots & \nonumber \\
R^A_{S_1^\pm} & \times & f_{\rm Central \, set}^{\rm CTEQ6.1M} \nonumber \\
R^A_{S_2^\pm} & \times & f_{\rm Central \, set}^{\rm CTEQ6.1M} \nonumber \\
& \vdots &, \nonumber
\end{eqnarray}
where the first one gives the central prediction, and others\footnote{The uncertainty sets in CTEQ6.1M and EPS09 correspond to similar 90\%-confidence criterion.} are pairwize contributing to the size of the upper and lower errors via Eq.~(\ref{eq:ASymmetricPDFerrors}).

We stress that this way of computing the uncertainty for d+Au cross-section is a simplification due to parametrizing the nuclear modifications on top of fixed proton baseline PDFs. For example, when the CTEQ error sets are multiplied by our central-set, the momentum is not strictly conserved. In other words, although the central prediction of our nPDFs is well-defined, the free and bound proton PDF analyses cannot be completely separated in the uncertainty analysis. The only way of obtaining completely self-consistent uncertainty predictions for d+Au collisions would require putting the free and bound proton PDF analyses together into a one single analysis in which the artificial separation to free proton uncertainties and those stemming from the nuclear modifications would not be required. In the absence of such master analysis, we recommend to calculate the uncertainties for absolute cross-sections as explained above.

\begin{figure}[htbp]
\center
\subfigure{
\includegraphics[scale=0.45]{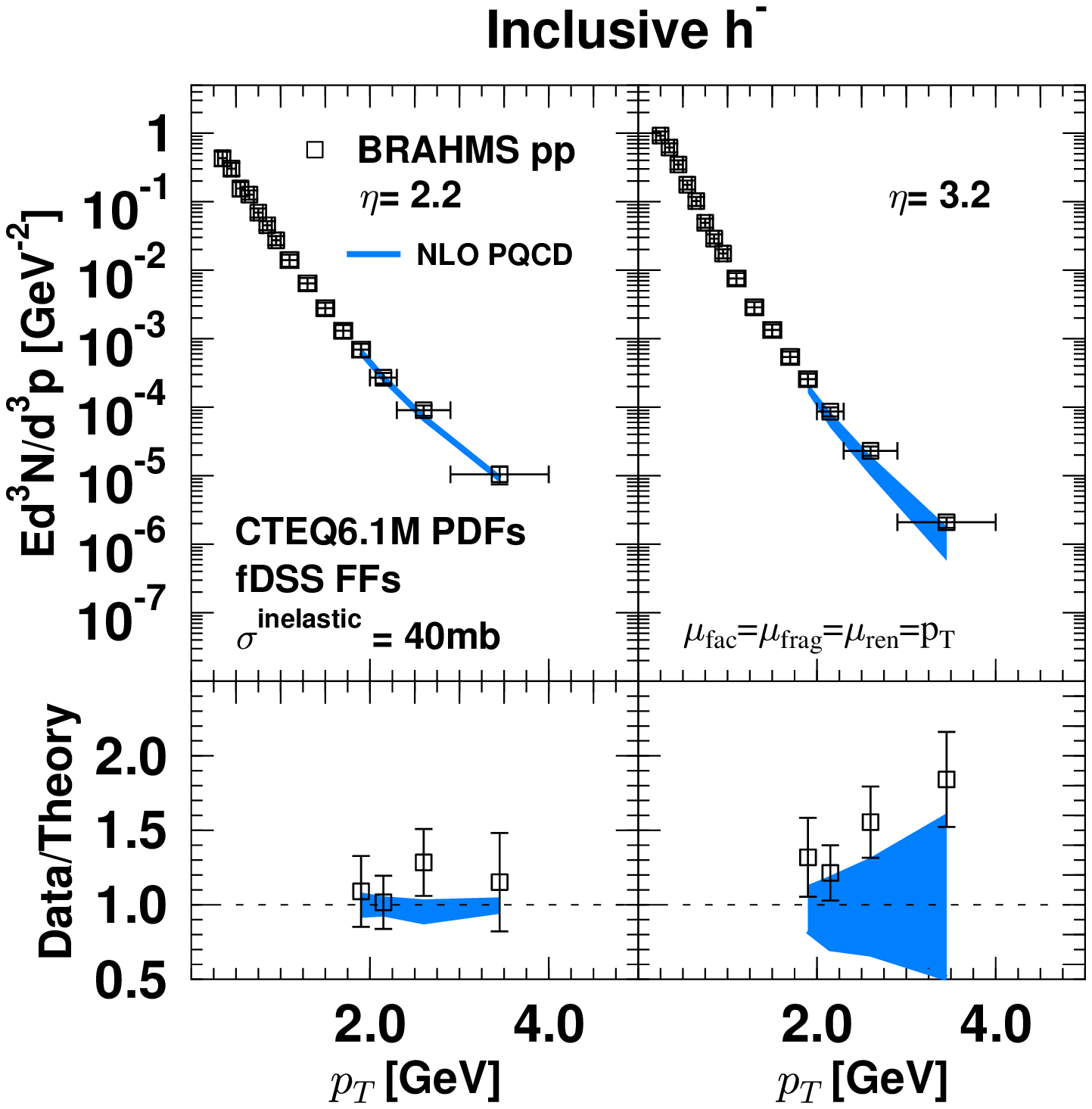}}
\subfigure{
\includegraphics[scale=0.45]{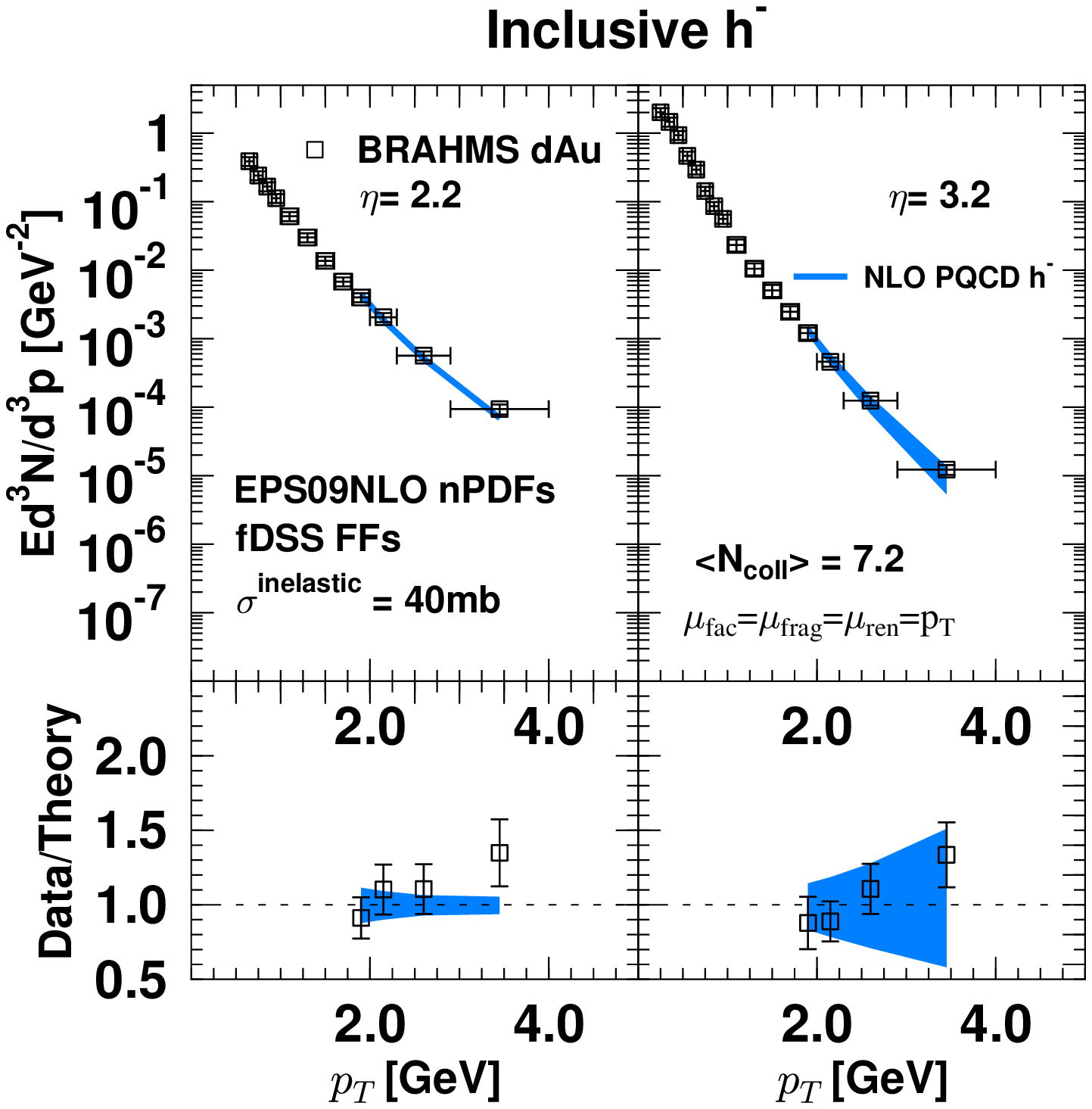}}
\caption[]{\small Inclusive yield of negatively charged hadrons $h^-$ in pp and dAu collisions. The experimental data shown by open squares is from \cite{Arsene:2004ux} with statistical and systematical errors added in quadrature, and the horizontal bars indicate the $p_T$ bin. The blue band indicates the 90\% confidence range derived from free proton and nPDF uncertainties. The calculated cross-sections have been averaged over the $p_T$-bin width.}
\label{Fig:BRAHMS1}
\end{figure}

The results and comparison with the BRAHMS data are shown in Fig.~\ref{Fig:BRAHMS1}, in which the blue bands indicate the PDF uncertainty range. The lower panels show the data/theory ratios.
Looking first at the $\eta = 2.2$ panels, the flatness of the data/theory ratio shows that the shapes of the measured distributions for both p+p and d+Au cases are well reproduced by the pQCD calculation. As these ratios are also very close to one, we conclude that the normalization factors $\sigma_{NN}^{\rm inelastic}$ and ${\langle N_{\rm coll}\rangle}$ are well estimated and that fixing the factorization scale to hadronic $p_T$ is also a valid choice.

\begin{figure}[htbp]
\center
\includegraphics[scale=0.7]{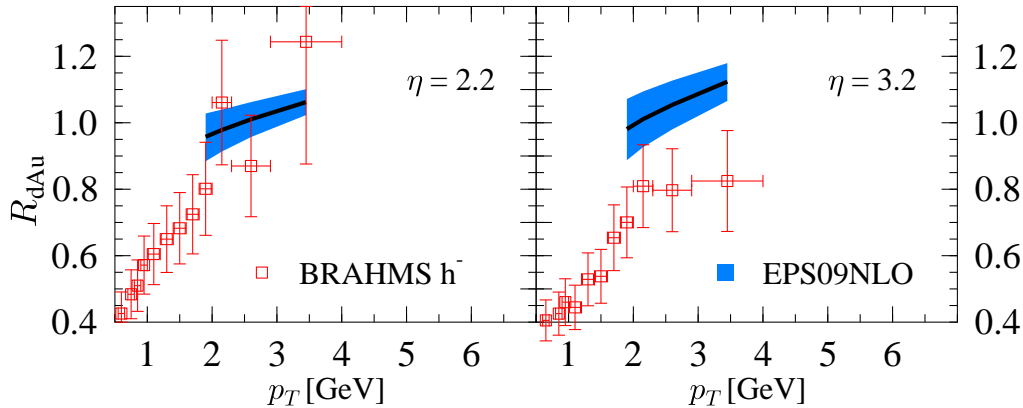}
\caption[]{\small The computed nuclear modification ratio $R_{\rm dAu}$ at forward rapidities (filled squares) for negatively-charged hadron production, compared with the BRAHMS  data \cite{Arsene:2004ux} (open squares). The error bars are the statistical and systematic uncertainties added in quadrature. The additional overall normalization uncertainty of $5\%$ is not shown. The blue bands are computed using the \texttt{EPS09NLO} sets.}
\label{Fig:BRAHMS2}
\end{figure}

In the $\eta = 3.2$ panels, one first observes that in the case of p+p collision the agreement between the theory and the data is clearly worse and the experimental and PDF errors are only barely overlapping. Such mismatch is not, however, observed in the d+Au cross-sections. Thus, we conclude that if the nuclear modification
\begin{equation}
R_{\rm dAu}  \equiv  \frac{1}{\langle N_{\rm coll}\rangle} \frac{d^2 N^{\rm dAu}/dp_T dy}{d^2 N^{\rm pp}/dp_T dy} \stackrel{\rm min. bias}{=} \frac{\frac{1}{2A} d^2\sigma^{\rm dAu}/dp_T dy}{d^2\sigma^{\rm pp}/dp_T dy}.
\end{equation}
is formed from these data, the result with $\eta=3.2$ will inevitably lie below the pQCD prediction. This is demonstrated in Fig.~\ref{Fig:BRAHMS2} where we show the comparison of the BRAHMS $R_{\rm dAu}$ data with the pQCD prediction from our global analysis. We stress that this happens not because we would incorrectly reproduce the d+Au cross-section but because the NLO pQCD calculation seems to undershoot the p+p data. This is the reason why we do not include these data to the current global nPDF analysis. However, as Fig.~\ref{Fig:BRAHMS1} shows, there is no reason why one should abandon the factorization for $h^-$ in d+Au collisions. Presumably, more statistics would be needed to resolve the discrepancy at $\eta=3.2$.

\section{Summary and Outlook}
\label{sec:Summary}

We have presented a global analysis of the nuclear corrections to the free proton PDFs at NLO accuracy in pQCD. Since the quality of the obtained fit is excellent and since there seems not
be any strong tensions between the fitted data sets\footnote{E.g. in EPS08 \cite{Eskola:2008ca} there was a significant tension between the DIS and forward BRAHMS data.}, we argue that our analysis supports the validity of collinear factorization in describing hard nuclear collisions such as DIS, DY and inclusive high-$p_T$ pion production. The qualities of both NLO and LO fits are almost identical, $\chi^2/N \approx 0.79$, but the NLO analysis somewhat reduces the inherent uncertainties. This is contrast to the free proton analyses where the quality of the NLO fit is significantly better (see e.g. \cite{Martin:2009iq}). In the near future, with new data for inclusive hadron and direct photon production from RHIC and LHC experiments, factorization will be tested further.

For the first time in the nuclear case, we have now also quantified the uncertainties originating from the experimental errors in a manner that we can release a collection of nPDF error sets. Using these 1+30 sets, the nPDF uncertainties can be propagated to any hard process of interest. In lieu of a still missing master analysis which would consistently combine both the free and bound proton PDFs, we recommend to estimate the uncertainties as follows: If an absolute cross-section $\sigma$ is considered the PDF and nPDF uncertainties should be combined independently, schematically as follows
$$
(\Delta \sigma)_{\rm Total}^2 = (\Delta \sigma)_{\rm Proton}^2 + (\Delta \sigma)_{\rm EPS09}^2.
$$
For the nuclear cross-section ratios $R$, --- like those in Eq.~(\ref{RF2RDY}) --- it should suffice to consider only the nPDF errors,
$$
(\Delta R)_{\rm Total}^2 = (\Delta R)_{\rm EPS09}^2
$$
(see Secs. \ref{sec:nPDFUncertainties} and \ref{sec:Application} for the details). In the latter case we do not recommend to include $(\Delta R)_{\rm Proton}^2$, as this would lead to an overestimation of the errors. Moreover, for consistency \texttt{EPS09} should be used in conjunction with the same free proton PDF set that was used in the analysis, namely CTEQ6.1M at NLO and CTEQ6L1 at LO.

The most up-to-date free proton PDF analyses (e.g. \cite{Martin:2009iq,Nadolsky:2008zw}) implements a somewhat better organized method for treating the heavy quarks than the ZM-VFNS employed in this analysis. Extension of the nPDF analysis to such general-mass scheme remains as a future work. Recently, the CTEQ collaboration has looked at the nuclear PDFs from a slightly different point of view by fitting the cross-section data from neutrino-Iron scattering \cite{Schienbein:2007fs}. Interestingly, the central fit points to a somewhat different nuclear modifications than what are obtained from electromagnetic DIS. Before making any strong conclusions about the possible difference, these data should be included in a global analysis and the possible tension between these and the other data sets carefully investigated. Also this is left as a future task.

\section*{Acknowledgements}
KJE and HP thank the Academy of Finland, Project 115262 and GRASPANP for financial support.
CAS is supported by Ministerio de Ciencia e Innovaci\'on of Spain under projects FPA2005-01963 and FPA2008-01177; by Xunta de Galicia (Programa Incite); and by the European Commission grant PERG02-GA-2007-224770. CAS is a Ram\'on y Cajal researcher.

\newpage
\begin{flushleft}
{\LARGE \bf Appendix}
\end{flushleft}

\appendix

\section{Error analysis}
\label{sec:ErrorAnalysis}

In this appendix we explain the error analysis in more detail. The basic formalism we present here was introduced in the original CTEQ articles \cite{Stump:2001gu,Pumplin:2001ct}, but we have taken guidelines also subsequent CTEQ and MRST papers e.g. \cite{Pumplin:2002vw,Martin:2002aw}.

\subsection{Hessian Matrix}
\label{sec:HessianMethod}

The Hessian approach is based on a quadratic expansion of $\chi^2 \equiv \chi^2(\{a\})$ near its minimum $\chi^2_0 \equiv \chi^2\{a^0\}$,
\begin{equation}
\chi^2 \approx \chi_0^2 + \sum_{ij} \delta a_i H_{ij} \delta a_j,
\label{eq:QuadraticApproximation}
\end{equation}
where $\delta a_i \equiv a_i-a_i^0$, and $H_{ij}$ is the Hessian matrix defined here by
\begin{equation}
H_{ij} \equiv \frac{1}{2} \frac{\partial^2 \chi^2}{\partial a_i \partial a_j}\Big|_{a=a^0}.
\end{equation}
Being symmetric, the Hessian matrix has a set of normalized eigenvectors ${\bf v}^{(k)}$ and eigenvalues $\epsilon_k$ such that
\begin{equation}
\sum_j H_{ij}v_j^{(k)} = \epsilon_k v_i^{(k)}, \qquad \sum_j v_j^{(k)}v_j^{(\ell)} = \delta_{k\ell}.
\end{equation}
Introducing a new set of coordinates $\{z\}$, as
\begin{equation}
\delta a_i = \sum_j v_i^{(j)} \sqrt{\frac{1}{\epsilon_j}} z_j,
\end{equation}
one finds that
\begin{equation}
z_k = \sqrt{\epsilon_k} \sum_j v_j^{(k)} \delta a_j,
\end{equation}
and
\begin{equation}
\chi^2 \approx \chi_0^2 + \sum_{i} z_i^2, 
\label{eq:Chi2inZ}
\end{equation}
which implies that in the region where the quadratic approximation (\ref{eq:QuadraticApproximation}) is valid, the new coordinates $\{z\}$ are independent --- the $\chi^2$ increases uniformly in an arbitrary direction in the ${z}$-space.

\subsection{Error propagation}
\label{sec:ErrorPropagation}

Let us consider any quantity $X$ that depends on the PDFs, that is $X \equiv X(\{z\})$. In a linear approximation, $X$ may be expanded in the vicinity of its central value $X_0 \equiv X(\{z=0\})$ as $X = X_0 + \Delta X$, where 
\begin{equation}
\Delta X \approx \sum_j \left( \frac{\partial X}{\partial z_j} \right) \delta z_j.
\end{equation}
Since $\chi^2$ grows uniformly in all $z$-space directions, the $z$-space vector that extremizes $\Delta X$ for a given $\Delta \chi^2$, is in the direction of the gradient of $X$ and has a length of $\sqrt{\Delta \chi^2}$. The components of this vector are
\begin{equation}
\delta z_i =  \sqrt{\Delta \chi^2} \left( \frac{\partial X}{\partial z_j} \right) {\left( \sum_j \left(\frac{\partial X}{\partial z_j}\right)^2 \right)^{-1/2}},
\end{equation}
giving
\begin{equation}
(\Delta X)^2_{\rm extremum} \approx \Delta \chi^2 \sum_j \left( \frac{\partial X}{\partial z_j} \right)^2. \label{eq:X_Extremum1}
\end{equation}
In order to facilitate the computation of the derivatives $\partial X/\partial z_k$ in Eq.~(\ref{eq:X_Extremum1}), we define auxiliary PDFs in several $z$-space coordinates:
\begin{eqnarray}
 S_{0} & = & (0,0,0, \ldots ,0) \nonumber \\
 S^{\pm}_1 & = & \pm \sqrt{\Delta \chi^2} \, (1,0,0, \ldots ,0) \nonumber \\
 S^{\pm}_2 & = & \pm \sqrt{\Delta \chi^2} \, (0,1,0, \ldots ,0) \nonumber \\
 & \vdots & \\
S^{\pm}_{N-1} & = & \pm \sqrt{\Delta \chi^2} \, (0,0, \ldots ,1,0) \nonumber \\ 
S^{\pm}_N & = & \pm \sqrt{\Delta \chi^2} \, (0,0, \ldots ,0,1) \nonumber.
\end{eqnarray}
The first set $S_0$ is the central set giving the minimum $\chi^2$, and in the other ones the fit parameters are changed by a fixed amount in each $z$-space direction separately. Using these PDF sets, the derivatives $\partial X/\partial z_k$ can be approximated by a finite difference
\begin{equation}
 \frac{\partial X}{\partial z_k} \approx \frac{X(S_k^\pm)-X(S_0)}{\pm \sqrt{\Delta \chi^2}} \approx \frac{X(S^+_k)-X(S^-_k)}{2\sqrt{\Delta \chi^2}},
\end{equation}
where $X(S_k^\pm)$ denotes the value of $X$ computed with the PDF set $S_k^\pm$. Upon insertion to Eq.~(\ref{eq:X_Extremum1}) we get
\begin{equation}
(\Delta X)^2_{\rm extremum} \approx \sum_k \left( X(S_k^\pm)-X(S_0) \right)^2 \approx \frac{1}{4} \sum_k \left( X(S^+_k)-X(S^-_k) \right)^2. \label{eq:X_Extremum2}
\end{equation}
These are the simplest equations by which the PDF uncertainties are propagated to any quantity $X$.

\begin{figure}[htbp]
\center
\includegraphics[scale=0.41]{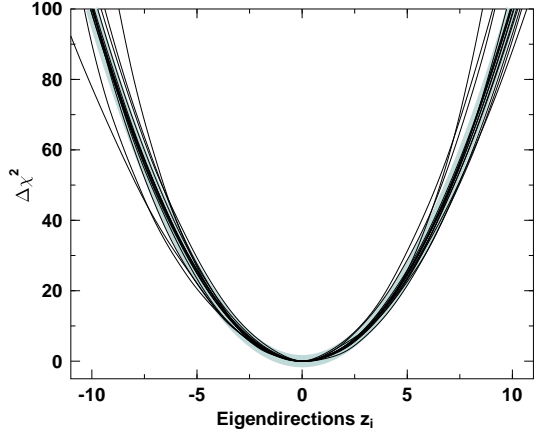}
\caption[]{\small This figure illustrates how the $\Delta \chi^2$ in our NLO analysis grows when the fit parameters are scanned along the eigenvector directions. The shaded light blue band indicates the ideal quadratic behaviour $\Delta \chi^2 = z_i^2$. We can see that the quadratic approximation is good but not perfect.}
\label{Fig:QuadraticTest}
\end{figure}

However, often the quadratic approximation is not perfect, revealed by non-ideal behaviour of $\chi^2$ in the $z$-space. From Fig.~\ref{Fig:QuadraticTest} such an behaviour can be concretely seen to some extent happen also in our NLO analysis. In such a case the growth of $\chi^2$ induced by $\sqrt{\Delta \chi^2}$ deviations in the eigenvector directions can be significantly different from the ideal $\Delta \chi^2$. Moreover, the eigendirections may exhibit very asymmetric behaviour, and the required deviations for $\chi^2$ to grow by $\Delta \chi^2$ may be significantly different on either side of the origin. This observation necessitates to modify the definition of the auxiliary PDFs to
\begin{eqnarray}
 S^{\pm}_1 & = & \pm \delta z_1^\pm \, (1,0,0, \ldots ,0) \nonumber \\
 S^{\pm}_2 & = & \pm \delta z_2^\pm \, (0,1,0, \ldots ,0) \label{eq:errorsets} \\
 & \vdots & \nonumber
\end{eqnarray} 
where $\delta z_i^+$ and $\delta z_i^-$ are the deviations in positive and negative side of $i$th eigendirection respectively, that cause $\chi^2$ to grow by $\Delta \chi^2$. 
Replacing the Eq.~(\ref{eq:X_Extremum2}) by the following prescription \cite{Nadolsky:2001yg}
\begin{eqnarray}
(\Delta X^+)^2 & \approx & \sum_k \left[ \max\left\{ X(S^+_k)-X(S^0), X(S^-_k)-X(S^0),0 \right\} \right]^2 \label{eq:X_Extremum3} \\
(\Delta X^-)^2 & \approx & \sum_k \left[ \max\left\{ X(S^0)-X(S^+_k), X(S^0)-X(S^-_k),0 \right\} \right]^2 \nonumber,
\end{eqnarray}
serves as a way to obtain the upper and lower uncertainties separately. If the quadratic approximation for $\chi^2$ and the linear approximation for $X$ were both perfect the Eq.~(\ref{eq:X_Extremum2}) and Eq.~(\ref{eq:X_Extremum3}) would lead to identical results.

\subsection{Choice of $\Delta \chi^2$?}
\label{sec:Choice_chi2}

In the derivation above, we assumed that there is a fixed $\Delta \chi^2$ which sets a $\chi^2$ boundary beyond which the global fit quickly deteriorates. In a statistically ideal case, when the uncertainties of the data are Gaussian-type and the theory is known to be ``right enough'', the one standard deviation corresponds to $\Delta \chi^2 = 1$ \cite{Amsler:2008zzb}. Such requirements can usually be met only if data from a very limited amount of sources are accepted --- like PDF fits including only HERA data \cite{Dittmar:2009ii}, where the ideal choice $\Delta \chi^2 = 1$ is indeed justified.

However, in a \emph{truly global} PDF analysis which, by definition, aims to combine data from as many independent experiments as possible, the situation becomes quite far from being ideal in the statistical sense: There are often sizable, unexplainable fluctuations within individual data sets and in practice all data sets are not mutually fully compatible. Thus, taking $\Delta \chi^2 = 1$ in such analysis becomes practically meaningless.

This is most obviously true also in the present analysis --- combining all data from the lightest nuclei up to the heaviest ones --- which poses an additional difficulty which is not met in the free-proton analysis: In lieu of large amount of data for each nucleus separately, we model the $A$-dependence of the fit parameters by a power law (\ref{eq:Adependence}). In this way we can evidently catch the correct general trend, but there will inevitably be some small pull between individual nuclei in various places. Thus, the choice for $\Delta \chi^2$ should account also for our inability to treat the individual nuclei separately.

Below, we first introduce somewhat extended statistical approach which is sometimes used in determining $\Delta \chi^2$, and continue with a physically better motivated one, which we use in our error analysis.
\\ \\
{\bf Statistically motivated method:}
\\ \\
Let us assume that the probability distribution for the global $\chi^2$-function is a simple Gaussian in the $N$-dimensional $z$-space
\begin{equation}
P_D(|{\bf z}|,N) = \frac{1}{\left( 2\pi \right)^{N/2}} e^{-{\bf z}^2/2}.
\end{equation}
The probability $P(\Delta \chi^2,N)$ for $\chi^2$ increasing at most $\Delta \chi^2$ is then obtained from
\begin{eqnarray}
P(\Delta \chi^2,N) & = & \int _{{\bf z}^2 \le \Delta \chi^2} \left( \prod_{i=1}^N dz_i \right) \, P_D(|{\bf z}|,N) =  \int _0 ^{\Delta \chi^2} dS \, P_D^{\chi^2}(N,S) \label{eq:idealchi2} \\
P_D^{\chi^2}(N,S) & = & \frac{1}{2\Gamma \left( N/2 \right)} \left(\frac{S}{2}\right)^{N/2-1} e^{-S/2}. \label{eq:chi2distribution}
\end{eqnarray}
For usual ``one standard deviation'' $P \approx 68.3 \%$ and for $N=15$ parameters, this approach would suggest $\Delta \chi^2 \approx 17$. This a way of choosing $\Delta \chi^2$ has been routinely employed e.g. by Kumano et al. \cite{Hirai:2007sx} and previously also  by us \cite{Eskola:2007my}.
\\ \\
{\bf Physically motivated method:}
\\ \\
An alternative way for finding $\Delta \chi^2$ is to examine how the $\chi^2$-contributions $\chi^2_k$ of individual data sets with $N_k$ data points, behave in the $z$-space (see e.g. \cite{Pumplin:2002vw}). Loosely speaking, by determining acceptable intervals for each of the $z$-space directions, and taking a suitable average, we can arrive at a physically better motivated value for $\Delta \chi^2$ than what is obtained from the idealized statistical way described above.

Restricting the range how much individual $\chi^2$-contributions $\chi^2_k$ are allowed to grow above the central set value $\chi^2_{0,k}$, is complicated by fluctuations in the data which can make $\chi^2_{0,k}/N_k \gg 1$, even if the overall agreement was visually good. In order to compensate for such effects and to treat all data sets equally, we follow the procedure suggested by the CTEQ and define normalized variables $\chi^2_{k,n}$ by
\begin{equation}
 \chi^2_{k} \rightarrow \chi^2_{k,n} \equiv \chi^2_{k} \left( \frac{N_k-2}{\chi^2_{0,k}} \right).
\end{equation}
Assuming that each $\chi^2_{k,n}$ follows the probability distribution of Eq.~(\ref{eq:chi2distribution}) with $N=N_k$, the most probable value of $\chi^2_{k,n}$ is attained by the central set $S_0$. For each experiment we denote the largest acceptable $\chi^2_{k,n}$ by $M(P,{\chi^2_{k}})$ and define it as a solution to equation
\begin{equation}
\int _0 ^{M(P,{\chi^2_k})} dS \, P_D^{\chi^2}(N_k,S) = P,
\end{equation}
for fixed confidence level $P$. Sliding each parameter $z_i$ across the parameter space and simultaneously monitoring the size of $\chi^2_{k,n}$, we determine the range
\begin{equation}
z^{(k)}_{i,-} \le z_i \le z^{(k)}_{i,+}
\end{equation}
that fulfills the criterion
\begin{equation}
\chi^2_{k,n} \le M(\chi^2_{k}).
\end{equation}
\begin{figure}[htbp]
\center
\includegraphics[scale=0.45]{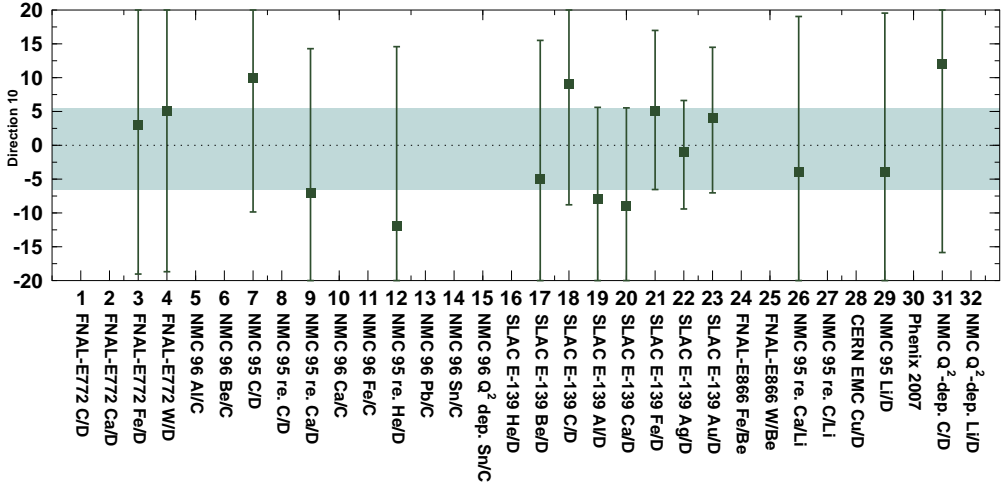}
\caption[]{\small The 90\% confidence limits for different data sets in the eigendirection direction $z_{10}$. The solid boxes indicate the locations of the minima for each $\chi^2_k$. No line is shown if the limits exceed the shown range. The global minimum is at $z=0$.}
\label{Fig:ConfidenceLimit}
\end{figure}

In Fig.~\ref{Fig:ConfidenceLimit}, we display the $P=0.90$ limits for eigendirection $z_{10}$ obtained by this procedure. The intersection $\cap_k [z^{(k)}_{i,-},z^{(k)}_{i,+}]$ gives us lower and upper bounds $z_i^-$ and $ z_i^+$, for which none of the individual $\chi^2$-contributions exceed their 90$\%$ confidence limits. This intersection is the shaded band in the Fig.~\ref{Fig:ConfidenceLimit}.

Finally, denoting the increase of $\chi^2$ in the boundaries by $\Delta \chi^2 (z_i^\pm)$, the following average is taken as an overall estimate for $\Delta \chi^2$:
\begin{equation}
 \Delta \chi^2 \equiv \sum_i \frac{\Delta \chi^2 (z_i^+) + \Delta \chi^2 (z_i^-)}{2N} \approx \sum_i \frac{(z_i^+)^2 + (z_i^-)^2}{2N},
\label{eq:finalDeltachi2}
\end{equation}
where the latter relation would be exact in an ideal case with perfectly quadratic $\chi^2$. Performing such a procedure we find $\Delta \chi^2 \approx 50$ --- three times the value that would be indicated by the idealized statistical method. 

We note that it is also possible to take $\delta z_i^\pm = z_i^\pm$ in Eq.~(\ref{eq:errorsets}). This choice is employed e.g. in a recent MSTW free proton analysis \cite{Martin:2009iq}. The benefit is that one circumvents the need for a single, ``master'' $\Delta \chi^2$ --- each direction $z_i$ is treated separately. We have tested also this approach and, in comparison to the method described above, observed somewhat larger uncertainty predictions in those $x$-regions, where there are no stringent data constraints. This is caused by larger parameter intervals allowed in those directions which do not quickly meet the 90\% confidence criterion, although the reproduction of several data sets gets worse, indicated by the approximate growth $\Delta \chi^2 \approx (\delta z_i)^2$. In other words, $\Delta \chi^2$ for the individual error sets would exhibit large differences among themselves. For this reason, we choose not to follow this precription, but construct the error sets in Eq.~(\ref{eq:errorsets}) such that correspond to a fixed increase of $\Delta \chi^2=50$.

\subsection{Calculation of the Hessian Matrix}
\label{sec:CalculationOfHessianMatrix}

A practical problem in applying the Hessian method is how to obtain a reliable approximation for the Hessian matrix. Ideally, if the $\chi^2$-function was smooth at the minimum and the $\Delta \chi^2$ very small, we would obtain a useful Hessian matrix as a side product of the {\ttfamily MINUIT} minimization. In reality, our $\chi^2$-function is a complicated functional of the fit parameters and due to the finite precision of the DGLAP solver and multiple numerical integrations, it generally becomes non-continuous  \cite{Pumplin:2000vx} in small parameter intervals. Such numerical noise makes it impossible to obtain a reliable Hessian matrix from a general purpose fitting program like {\ttfamily MINUIT}.

\begin{figure}[htbp]
\center
\subfigure[Uncorrelated]{
\includegraphics[scale=0.5]{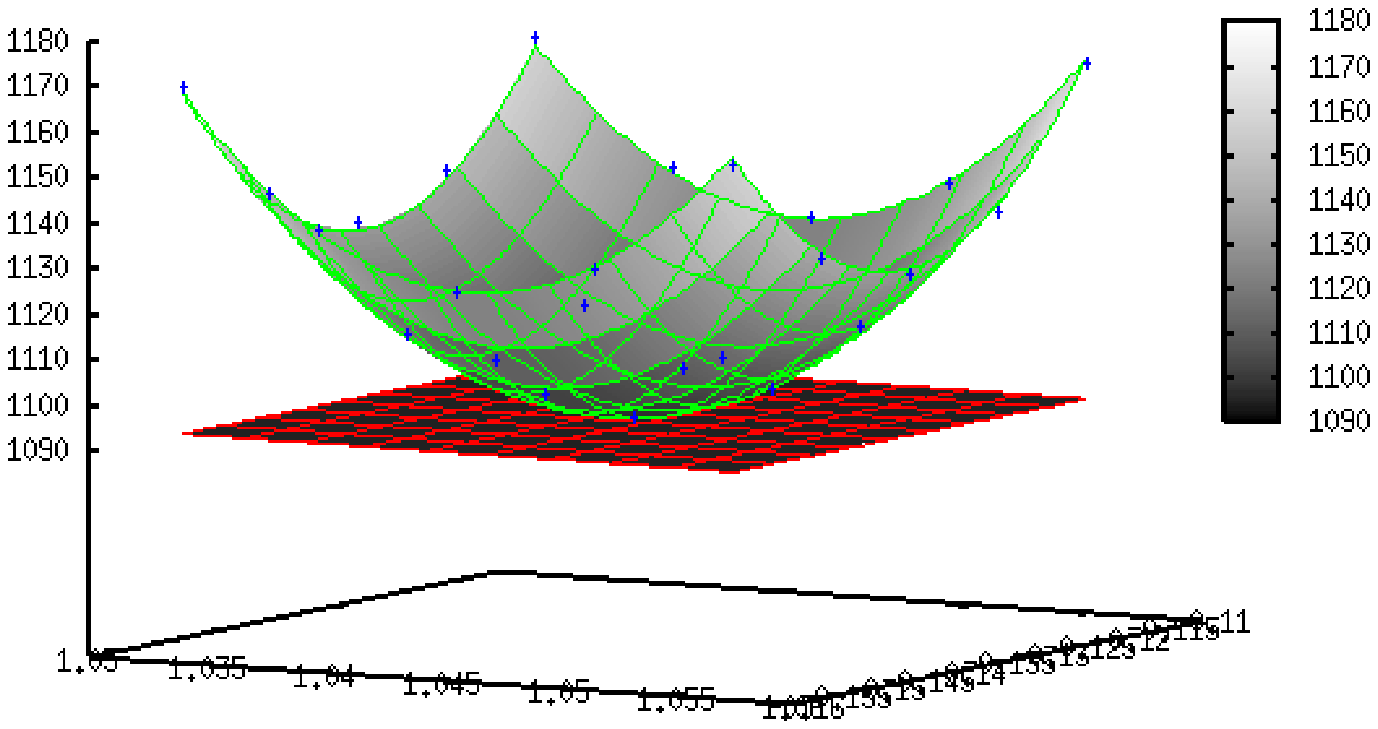}}
\subfigure[Correlated]{
\includegraphics[scale=0.5]{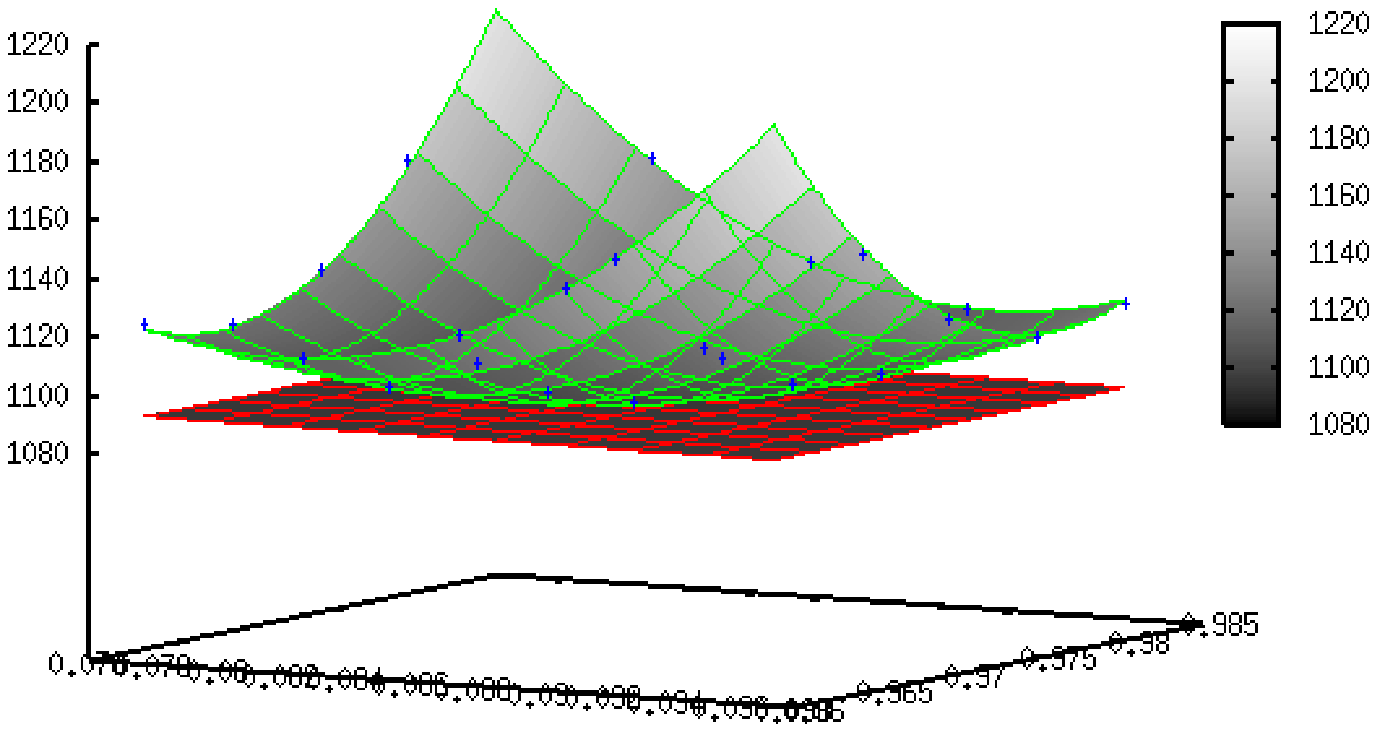}}
\caption[]{\small Two illustrative examples how our $\chi^2$-function may look in a two-parameter plane. The individual blue points are the exact $\chi^2$ values and the shaded surface results from the fit of Eq.~(\ref{eq:chi2fit}). In the figure (a), the two parameters are fairly uncorrelated and the $\chi^2$ increases almost uniformly in all directions. In the other case (b) there is a significant correlation between the two parameters revealed by the shallow ``valley'' going from a corner to another.}
\label{Fig:HessianFit}
\end{figure}

In order to obtain an estimate for the Hessian matrix, we adopt the following algorithm:
\begin{enumerate}
\item
Determine the parameter step sizes $\pm \delta a_i$ from their central values $a_i^0$ that give an approximately fixed increase in $\chi^2$.
\item
Compute the $\chi^2$ values for each parameter pair in a two dimensional grid spanning the range  $(a_i,a_j) \in [a_i^0 - \delta a_i,a_i^0 + \delta a_i] \times [a_j^0 - \delta a_j,a_j^0 + \delta a_j]$.
\item
Fit the coefficients with quadratic function of the form
\begin{equation}
f(a_i,a_j) = \chi^2_0 + c_i(a_i-a_i^0)^2 + c_j(a_j-a_j^0)^2 + c_{ij}(a_i-a_i^0)(a_j-a_j^0)
\label{eq:chi2fit}
\end{equation}
to the computed $\chi^2$-values at the grid points, and read off the Hessian matrix entries: $H_{ij} = c_{ij}/2$. The diagonal elements $H_{ii}$ are obtained by a similar fit but in one dimension.
\end{enumerate}
In Fig.~\ref{Fig:HessianFit} we show two examples of how the $\chi^2$ typically looks as a function of two parameters. The shaded surface shows what is obtained from the fitted function of Eq.~(\ref{eq:chi2fit}). From figures like this, we have verified how well the quadratic function (\ref{eq:chi2fit}) fits the neighborhood of the $\chi^2_0$ well. These figures clearly reveal the two-parameter correlations too.

Obtaining a trustworthy approximation for the Hessian matrix from the above fitting method, requires knowledge of the desired $\Delta \chi^2$ beforehand for adjusting the parameter intervals to be physically relevant: If too small or large step sizes are used to construct the Hessian matrix, the resulting quadratic approximation (\ref{eq:QuadraticApproximation}) may be bad for the $\Delta \chi^2$ which will eventually be required by Eq.~(\ref{eq:finalDeltachi2}). Therefore also this procedure is necessary somewhat iterative: To begin with, the Hessian matrix is computed by fixing the parameter step sizes to cause a fixed increase in $\chi^2$, say 25. Following the procedure outlined in earlier Sec.~\ref{sec:Choice_chi2}, an appropriate $\Delta \chi^2$ is then determined. Inspecting the behaviour of $\chi^2$ as a function of the deviations in each eigendirections, as shown in Fig.~\ref{Fig:QuadraticTest}, one may judge whether the achieved quadratic approximation is valid or not near the obtained $\Delta \chi^2$. If needed, one may then try to adjust the parameter intervals in the computation of the Hessian accordingly, in order to improve the quadratic approximation.

\section{How to speed up computation of pion production}
\label{InclusivePionProduction}

In this appendix we sketch the method to speed up the numerical calculation of inclusive pion production cross-section in ${\rm d} + {\rm Au}$ collisions

\begin{equation}
{\rm d}(K_1) + {\rm Au}(K_2) \rightarrow {\rm \pi}(K_3) + X \nonumber,
\end{equation}
where $K_1$ and $K_2$ denote the incoming momenta, and $K_3$ the momentum of the observed pion.  The cross-section can be written as a convolution of the PDFs $f$, the fragmentation function $D_{l \rightarrow \pi}$ for parton $l$ to make a pion, and the partonic cross-section $\hat{\sigma}$ \cite{Aversa:1988vb}
\begin{eqnarray}
\label{eq:1sthadroncrosssection}
E_3 \frac{d\sigma({\rm d} + {\rm Au} \rightarrow \pi + X)}{d^3K_3}  & = & \sum _{ijl} \int dx_1 \int dx_2 \int \frac{dx_3}{x_3^2} f_i^{\rm d}(x_1,\mu^2_{\rm fact}) f_j^{\rm Au}(x_2,\mu^2_{\rm fact}) \\
 & & D_{l \rightarrow \pi}(x_3,\mu^2_{\rm frag}) \, p_3^0 \frac{d\hat{\sigma}(p_1^i + p_2^j \rightarrow p_3^l, \mu^2_{\rm ren})}{d^3p_3}
_{
\begin{array}{l}
\vline \; p_1 = x_1K_1 \\
\vline \; p_2 = x_2K_2 \\
\vline \; p_3 = K_3/x_3 \\
\end{array}
}  \nonumber.
\end{eqnarray}
Here one faces a practical problem of significantly extended computing time consumed in evaluating the triple integrals, apparently making the fitting procedure --- involving tens of thousands of calls --- very slow. However, if the factorization scale $\mu_{\rm fact}^2$ is chosen not to depend on the partonic momenta, e.g. fixing all scales to a common scale $\mu^2$ proportional to the transverse momentum of the pion, one can define a set of weighting functions
\begin{equation}
F_j(x_2,\mu^2) =  \sum _{il} \int dx_1 \int \frac{dx_3}{x_3^2} f_i^{\rm d}(x_1,\mu^2) D_{l \rightarrow h}(x_3,\mu^2) \, p_3^0 \frac{ d\hat{\sigma}(p_1^i + p_2^j \rightarrow p_3^l,\mu^2)}{d^3p_3},
\end{equation}
which can be evaluated and tabulated in advance as $F_j(x_2)$ do not depend on the nuclear PDFs. After such preparations, the Eq.~(\ref{eq:1sthadroncrosssection}) can be evaluated in advance up a to single integral
\begin{equation}
E_3 \frac{d\sigma}{d^3K_3}  =  \sum_j \int dx_2 F_j(x_2,\mu^2) f_j^{\rm Au}(x_2,\mu^2). \label{eq:2ndhadroncrosssection}
\end{equation}
With the above trick we overcome the problem of slowness and make the inclusion of hadron production data feasible also in NLO. We should also note that this method is possible as we take the free proton (and deuterium) PDFs as fully known, i.e these are not subject to iteration here. If this were not the case, only one integral could be predone in Eq.~(\ref{eq:1sthadroncrosssection}).

\end{document}